\documentclass[12pt]{article}
\usepackage{graphicx}
\usepackage{url}
\usepackage{amssymb}
\usepackage{cite}
\usepackage[margin = 1.25in]{geometry}
\usepackage[colorlinks = true, linkcolor = blue, urlcolor = blue,
      citecolor = blue, anchorcolor = blue]{hyperref}

\newcommand\pubnumber{SLAC--PUB--17356}
\newcommand\pubdate{\hfill November, 2018}

\def\SLAC{SLAC,
    Stanford University, Menlo Park, California 94025 USA}
\def\doeack{\footnote{Work supported by the US Department of Energy,
                     contract DE--AC02--76SF00515.}}

\def\Title#1{\begin{center} {\Large #1 } \end{center}}
\def\Author#1{\begin{center}{ \sc #1} \end{center}}
\def\Address#1{\begin{center}{ \it #1} \end{center}}

\def\submit#1{\begin{center}Submitted to {\sl #1} \end{center}}
\newcommand\pubblock{\rightline{\begin{tabular}{l} \pubnumber\\
         \pubdate \end{tabular}}}
\newenvironment{Abstract}{\begin{quotation} \begin{center}
                       ABSTRACT
     \end{center}\bigskip  }{\end{quotation}}

\def\submit#1{\begin{center}Submitted to {\sl #1} \end{center}}
\def\Acknowledgements{\bigskip  \bigskip \begin{center} \begin{large}
             \bf ACKNOWLEDGEMENTS \end{large}\end{center}}

\def\beq{\begin{equation}}
\def\eeq#1{\label{#1}\end{equation}}
\def\eeqn{\end{equation}}

\newenvironment{Eqnarray}%
   {\arraycolsep 0.14em\begin{eqnarray}}{\end{eqnarray}}
\def\beqa{\begin{Eqnarray}}
\def\eeqa#1{\label{#1}\end{Eqnarray}}
\def\eeqan{\end{Eqnarray}}
\def\CR{\nonumber \\ }

\def\leqn#1{(\ref{#1})}

\let\bar=\overbar

\def\eg{{\it e.g.}}

\def\VEV#1{\left\langle{ #1} \right\rangle}
\def\bra#1{\left\langle{ #1} \right|}
\def\ket#1{\left| {#1} \right\rangle}

\def\lsim{\mathrel{\raise.3ex\hbox{$<$\kern-.75em\lower1ex\hbox{$\sim$}}}}
\def\gsim{\mathrel{\raise.3ex\hbox{$>$\kern-.75em\lower1ex\hbox{$\sim$}}}}

\def\M{{\cal M}}
\def\O{{\cal O}}

\def\half{\frac{1}{2}}

\def\del{\partial}
\def\Dslash{\not{\hbox{\kern-4pt $D$}}}
\def\dslash{\not{\hbox{\kern-2pt $\del$}}}

\def\ee{e^+e^-}

\def\msb{{\bar{\scriptsize M \kern -1pt S}}}

\def\drb{{\bar{\scriptsize D \kern -1pt R}}}

\makeatletter
\def\section{\@startsection{section}{0}{\z@}{5.5ex plus .5ex minus
 1.5ex}{2.3ex plus .2ex}{\large\bf}}
\def\subsection{\@startsection{subsection}{1}{\z@}{3.5ex plus .5ex minus
 1.5ex}{1.3ex plus .2ex}{\normalsize\bf}}
\def\subsubsection{\@startsection{subsubsection}{2}{\z@}{-3.5ex plus
-1ex minus  -.2ex}{2.3ex plus .2ex}{\normalsize\sl}}

\renewcommand{\@makecaption}[2]{%
   \vskip 10pt
   \setbox\@tempboxa\hbox{\small #1: #2}
   \ifdim \wd\@tempboxa >\hsize     
       \small #1: #2\par          
     \else                        
       \hbox to\hsize{\hfil\box\@tempboxa\hfil}
   \fi}

\makeatother

\def\A{{\cal A}}
 
\def\G{{\cal G}}
\def\F{{\cal F}}

\def\C{{\cal C}}

\def\Z{{\cal Z}} 

\begin{document}
\begin{titlepage}
\pubblock

\vfill
\Title{Fermion Pair Production in $SO(5)\times U(1)$ Gauge-Higgs
  Unification Models}
\vfill
\Author{Jongmin Yoon and Michael E. Peskin\doeack}
\Address{\SLAC}
\vfill
\begin{Abstract}
We compute fermion pair production cross sections in $\ee$
annihilation in models of electroweak symmetry breaking in warped
5-dimensional space.  Our analysis is based on a framework with no 
elementary scalars, only gauge bosons and fermions in the bulk. We
apply a novel Green's function method to obtain an analytic
understanding of the new physics effects across the parameter space.
We present results for $\ee\to b\bar b$ and $\ee\to t\bar t$.  The
predicted effects will be visible in precision measurements.
Different models of $b$ quark mass generation can be distinguished by
these measurements already at 250~GeV in the center of mass. 
\end{Abstract}
\vfill
\submit{Physical Review D}
\vfill
\end{titlepage}

\hbox to \hsize{\null}

\newpage

\tableofcontents

\def\thefootnote{\fnsymbol{footnote}}
\newpage
\setcounter{page}{1}

\setcounter{footnote}{0}

\section{Introduction}

The Standard Model (SM)  gives an excellent description of physics
at the energies that we now explore at accelerators. But still it is a
compelling idea that the SM is incomplete.  If there is an
explanation for the electroweak symmetry breaking of the SM
and the magnitudes of the quark and lepton masses, this requires
a generalization of the SM with new fundamental
interactions associated with the Higgs field. 

Many extensions of the SM have been explored
theoretically.   However, it is our opinion that there is still much to
learn about models in which the
Higgs field is composite~\cite{Csakireview,CGT}.
   In a series of papers \cite{YPone,YPtwo}, we have attempted
to gain new insight into this class of models by considering these
models in a Randall-Sundrum framework~\cite{RS}  in which the SM is 
extended to  5-dimensional anti-de Sitter space.  The  gauge group of
the SM can be enlarged
so that the Higgs doublet can appear as the 5th components
of gauge fields~\cite{gaugeHiggsone,gaugeHiggstwo}.   In this context,
it is
 possible to compute the
Higgs potential and find a dynamical origin for electroweak symmetry
breaking \cite{CNP}.

 In \cite{YPtwo} we examined in some detail a
particular model based on $SO(5)\times U(1)$ gauge symmetry in the
5-dimensional bulk.   This follows a path defined about ten  years ago 
by Agashe, Contino, and Pomarol~\cite{ACP}.   By making use of some
new model-building
strategies, we were able to find a class of models with an adjustible 
hierarchy between the electroweak scale and the mass scale of the
intrinsically 5D  dynamics.  Specifically, this ``little hierarchy''
is controlled by a mixing angle parametrized by $s = \sin\theta$.  
The little hierarchy must be large to avoid constraints from 
precision electroweak measurements.  Then  $s$ is small and can 
 be used as an 
expansion parameter. The predictions of the model can be expressed
as corrections to SM formulae of order $s^2$ and higher,
and this presentation gives useful insight into their structure.

In this paper, we continue this study by describing the implications
of measurements of quark and lepton pair production.
For definiteness, we consider in detail the reaction $\ee\to b\bar b$, where
the electron is assumed to be structureless, that is, confined to the
extreme UV region of the 5D Randall-Sundrum space.   This physics  has
been explored previously by Funatsu, Hatanaka, Hosotani, and Orikasa
in \cite{Hosotani}.   We show here that there is not a
unique prediction for the modification  of the $\ee\to  b\bar b$ cross
section, but, rather, a discrete set of predictions depending on the scheme for
producing the mass of the $b$ quark.

In \cite{YPtwo}, we generated the mass of the top quark through 
dynamical symmetry breaking.   In that paper, we left open the
question of how the lighter quarks and leptons receive their masses.
These must arise by feed-down from the top quark mass.  The flavor
mixing should occur at high energies, above the scale at which the top
quark mass is generated.  According to
the dictionary linking the 5D and 4D descriptions of the theory,  this mixing
involving the light fermions should be represented by a boundary 
condition on the UV
boundary of the Randall-Sundrum space.   Here, we will describe
several  distinct scenarios for this flavor mixing and show that
these are reflected in different predictions for observable $\ee\to
b\bar b$
cross
sections.  Typically, the effects become large enough to be observed
at 500~GeV in the center of mass, but in some cases we find observable
deviations from the SM at 250~GeV.

 Our analysis
has a straightforward extension to $\ee\to t\bar t$.   Corrections to
the $t\bar t$ cross section in RS models have been studied in numerous
references, reviewed in \cite{Richard}, and additional model
predictions 
been presented more recently in \cite{deCurtis}.   We believe that our analytic approach to
the parameter space of RS models contains new insight into the size
and nature of  
the expected effects.

The outline of this paper is as follows:   In Section~2, we review
well-known results on $\ee\to f\bar f$ for massless fermions with
unbroken $U(1)$ gauge symmetry (Randall-Sundrum QED)
\cite{Gherghetta}, and show how these follow from the formalism of
\cite{YPtwo}. 
In Section~3, we review the 
features of the 5D $SO(5)\times U(1)$ RS model that are
essential for our discussion here.  In Section~4, we present the 
structure of the electroweak boson propagators in this model. 
 In
Section 5, we describe schemes for generating the mass of the $b$
quark in our model and work out the implications of each for the
helicity-dependent cross sections in $\ee\to b\bar b$.  Precision
measurements of $\ee\to b\bar b$ at an $\ee$ collider, even at
250~GeV, can distinguish these models.   In Section 6,
we
generalize this analysis to the computation of the
helicity-dependent cross sections for $\ee\to t\bar t$ and present our
numerical predictions.  The resulting effects are large enough to 
be readily discovered in precision $\ee$ collider
experiments.  Section 7  gives our conclusions.

\section{Randall-Sundrum QED}

In this section we review results on the process $\ee\to f\bar f$
mediated by a $U(1)$ bulk gauge field in the Randall-Sundrum space. We
begin our discussion with the basic formalism for RS geometry and 5D
bulk fields.
 The notation follows that of \cite{YPone} and \cite{YPtwo}.

\subsection{Overview}

We consider a model of gauge and fermion fields living in the interior
of a slice 
of 5-dimensional anti-de Sitter space 
\beq
      ds^2 =  {1\over (kz)^2} [ dx^m dx_m - dz^2 ] 
\eeq{metric}
 with nontrivial boundary conditions at $z = z_0$ and $z = z_R$, with
 $z_0 < z_R$. 
Then $z_0$ gives the position of the ``UV brane'' and
$z_R$ gives the position of the ``IR brane''.     The perhaps more
physical metric
\beq
     ds^2 = e^{-2k x^5} dx^m dx_m  -   (dx^5)^2
\eeq{RSmetric}
is related by $kz = \exp[ k x^5]$.  We
take the  size of the interval in $x^5$ to be $\pi R$.  Then
\beq
        z_0 = {1/k}  \qquad       z_R  = 1/k_R \equiv    e^{\pi k R}/k \ . 
\eeq{zzeroRdef}
The scales $k$ and $k_R$ set the ultraviolet and infrared boundaries
of the dynamics described by the 5D fields.

The bulk action of gauge fields and fermions in RS  is 
\beq 
S_{bulk} = \int d^4 x dz\ \sqrt{-g}  \biggl[ - {1\over 4} g^{MP} g^{NQ}
  F^a_{MN}
F^a_{PQ} + \bar \Psi  \bigl[ i e^M_A \gamma^A {\cal D}_M 
- m_{\Psi} \bigr] \Psi  \biggr]    \ ,
\eeq{bulkaction}
We will notate gauge fields as $A_M^A$, where $M = 0,1,2,3,5$, with
lower case $m = 0,1,2,3$. Fermion fields are 4-component Dirac
fields. Here, we will typically break these up into 2-component
spinors with negative and positive 4D chirality: 
 $\Psi = (\psi_L, \psi_R)$. We will parametrize the 5D Dirac mass
 using the dimensionless parameter $c = m_{\Psi}/k$. In
 principle,  this parameter is different for each gauge multiplet of fermion fields.
 In our formalism, the Higgs field is a background gauge
field, so we will quantize in the Feynman-Randall-Schwartz  background
field gauge~\cite{RandallSchwartz}. 

 For concreteness, we  will 
be interested in values of $k_R$ of order 1~TeV and values of $k$ of
order 100~TeV.  Thus, we imagine that $z_0$ is at a flavor dynamics scale
rather than at the Planck scale.   Still, it will be accurate to
ignore terms of order  $z_0^2/z_R^2$, and we will do so throughout our
calculations.  We will not ignore terms of order
$\log(z_R/z_0)$ or, more generally, terms of order
$(z_0/z_R)^{c-1/2}$.

The physics beyond the UV cutoff scale $k$ can affect the dynamics of
the 5D theory. 
To model this, we apply nontrivial boundary conditions on the UV brane.  First, we
will allow mixing of fermions that belong to different gauge
multiplets in the bulk but have the same quantum numbers under the
SM subgroup of the bulk gauge group.  Second, we include 
localized 
kinetic terms on the UV brane  for the gauge fields and fermions:
\beq 
S_{UV} = \int d^4 x dz \ \biggl(\sqrt{-g} z_0 \delta(z-z_0) \biggr) 
\biggl[-{1 \over 4} a_B g^{mp} g^{nq} F_{mn} F_{pq} + 
a_\psi \psi_L^\dagger i (kz \bar\sigma^m) D_m \psi_L\biggr] \  ,
\eeq{UVaction}
where $a_B$, $a_\psi$ are constant parameters. 
 A similiar term for $\psi_R$ is also allowed but will not be used
 here.  In either case, for the fermions, 
 the boundary term has a substantial effect only for a field that has
  a UV-localized zero-mode. The formalism of the 
boundary kinetic terms is discussed in detail in  \cite{YPtwo}. 

We will work in Fourier space for the four extended dimensions and in
coordinate space for the 5th, warped,  dimension.
The solutions of field equations in the RS geometry with fixed   Minkowski
momenta  are then given in terms of 
Bessel functions in the form~\cite{GW,DHR,GP}
\beq
     \Phi =   z^a [ A  J_\nu(pz) + B Y_\nu(pz) ] e^{-ip\cdot x} \ . 
\eeq{Besselsolutions}
It is useful to define  combinations of the Bessel functions so that
the solutions \leqn{Besselsolutions}, as a function of $z = z_1$,  have 
definite
 boundary conditions at a point $z  = z_2$.   Thus we set
\beq
    G_{\alpha\beta}(z_1,z_2) = {\pi \over 2} \left[ J_{\alpha}(p z_1) 
 Y_{\beta} (pz_2)
   -  Y_\alpha(p z_1)  J_\beta (pz_2) \right] \ , 
\eeq{Gdef}
where $\alpha, \beta = \pm 1$.  For solutions to the Dirac equation, 
the orders of the Bessel functions
depend on the parameter $c$ according to 
\beq
  \mbox{for} \ \alpha,\beta =+1\  : \ \nu_+ =  c+\half \ ; \qquad 
 \mbox{for} \ \alpha,\beta =-1\  : \  \nu_- = c-\half    \ .
\eeqn
For gauge fields,  the same combinations of Bessel functions apply
with  $c = 1/2$, with $\alpha = +1$ giving the solutions for $A_m$ and
$\alpha = -1$ giving the solutions for $A_5$.   For gauge fields $A_m$,
 $G_{+-}(z,z_R)$, $G_{++}(z,z_R)$ give
 solutions with 
Neumann  and Dirichlet boundary conditions, respectively, at $z =
z_R$.   For fermions $\psi_L$, $G_{+-}(z,z_R)$, $G_{++}(z,z_R)$
give solutions with $\psi_R = 0$ and $\psi_L = 0$, respectively, at $z
= z_R$.  In the rest of paper, we will denote these
 boundary conditions by $+$ or $-$. For example, $(+-)$ represents
 Neumann or $\psi_R = 0$ boundary condition on the UV brane and
 Dirichlet or $\psi_L = 0$ boundary condition on the IR brane.
Further discussion of  our formalism and additional 
properties of the Green's functions $G$ are given in the appendices of
\cite{YPtwo}.

In general, in this paper, when a $G$ function appears without
arguments, it is 
\beq
            G_{\alpha\beta} \equiv  G_{\alpha\beta}(z_0,z_R)  \ . 
\eeq{Gnoarg}
For fields with a boundary kinetic term with coefficient $a_B$, we will
find it convenient to define
\beq
           G_{B- \pm } (z, z_R) \equiv    G_{-\pm }(z, z_R) + a_B p
           z_0 G_{+ \pm}(z, z_R) \ . 
\eeq{GBnotation}
We define further
\beq
              L_B  \equiv  G_{B--} (z_0, z_R)\biggr|_{p = 0} \ . 
\eeq{LBdefin}
For $c = 1/2$, the case of a bulk gauge field,  $L_B = \log (z_R/z_0)
+ a_B$ and gives the relation between the 5D coupling constant and the
(dimensionless) 4D coupling constant, 
\beq
           g^2 =  {  g_5^2 k\over L_B }  =    {g_5^2 k \over
             \log(z_R/z_0) + a_B } \ . 
\eeq{gfourfive}

\subsection{Chiral fermion wavefunctions}

Massive fermions in 5D have nonzero components for  both $\psi_L$ and
$\psi_R$.  However, with appropriate boundary conditions, 5D fermion
fields can have
zero modes that can be interpreted as chiral quarks and leptons
\cite{GP,Grossman}. 

A left-handed fermion zero mode has the form
\beq
\psi_L = f_L(z) \, u_L(p) \, e^{-ip\cdot x} \qquad \psi_R = 0 \ ,
\eeq{genleftzero}
where $u_L(p)$ is the usual 2-component massless spinor. 
Solving the RS Dirac equation, we
find that the zero mode has the form
\beq
f_L(z) = \left[ { 1-(z_0/z_R)^{1-2c} \over 1-2c} + a_\psi \left(z_0
    \over z_R\right)^{1-2c} \right]^{-1/2} {k^2 \over z_R^{\half-c} }
\  z^{2-c}  \ .
\eeq{leftzero}
We have normalized the zero mode wavefunction so that 
\beqa
   \int dz \sqrt{g} (1 + a_\psi z_0 \delta(z-z_0)) \bar \Psi (kz
   \gamma^0) \Psi & & \CR & & \hskip -1.6in
 =  \int dz { kz \over
     (kz)^5} (1 + a_\psi z_0 \delta(z-z_0)) \bigl| f_L(z) \bigr|^2  = 1 \, .
\eeqa{leftnorm}
Notice that, according to \leqn{leftzero} or \leqn{leftnorm}, a finite
contribution to the normalization integral is contained in a singular
piece of the wavefunction at $z = z_0$.  Away from this point, the
fermion wavefunction  is smooth.   In the following sections, we will
need to compute expectation values in the 5D fermion wavefunctions.  It is
convenient to use the formula
\beq
 \VEV{g(z)}  = \int^{z_R}_{z_0} {dz \over (kz)^4} \bigl| f_L(z) |^2  (g(z) - g(z_0))
 +  g(z_0) \ . 
\eeq{expformforf}
This avoids explicit consideration of the singularity at $z= z_0$.
Some further discussion of this point is given in Appendix A. 

A right-handed fermion zero mode has the form 
\beq
\psi_R = f_R(z) \, u_R(p) \, e^{-ip\cdot x} \qquad \psi_L = 0 \ ,
\eeq{genrightzero}
where $u_R(p)$ is the usual 2-component right-handed massless spinor.
The function $f_R(z)$ is obtained from $f_L(z)$ by sending $c \to -c$
and, for our choice of boundary conditions, setting $a_\psi = 0$.

For $ c \gg 1/2$, $f_L(z)$ is localized near  the UV brane. That is,  we can
consider the wave-function of the UV-localized zero mode as well
approximated by  delta function at $z_0$:
\beq
{ \bigl| f_L \bigr|^2 \over (kz)^4 } \rightarrow \delta(z-z_0), \quad \textrm{for} \ c \gg \half \ .
\eeq{UVlimit}
For a left-handed zero mode, the boundary kinetic term $a_\psi$ can
make a substantial effect only if $c \gsim 1/2$. In this case,
$a_\psi$ shifts the $\psi_L$ wavefunction further towards the UV boundary. For $c
\gsim 1/2$, the right-handed zero mode is localized near the IR brane
and therefore its boundary term would have negligible effect. It is
therefore justified to 
ignore possible  boundary term for IR-localized zero-modes. 

Eventually, the zero-mode fermions will obtain mass by mixing of a
left-handed zero mode with a right-handed zero mode.  We will discuss
schemes for mass generation for the $b$ quark in Sections~5.   However, if we
generate masses that are small compared to $m_Z$, it will always be a
good approximation to neglect the direct effects of the masses and
mixings in the computation of cross sections.  Since the $A_5$
components of gauge bosons have matrix elements only between $\psi_L$
and $\psi_R$ components of fermion fields, we will then also neglect
$A_5$ exchange.

\subsection{$\ee \to f \bar f$ in RS}

Using the Feynman-Randall-Schwartz gauge fixing with $\xi = 1$, we can 
compute the Green's functions for gauge fields
\beq
\VEV{ \A_m(z_1,p) \A_n(z_2,-p)} = \eta_{mn} \G(z_1,z_2,p) \ .
\eeqn
The details of the Green's function computation can be found in the
appedix of \cite{YPtwo}. For a non-Abelian bulk gauge group,
$\G(z_1,z_2,p)$ will be a matrix in the group indices.

Combining this with the description of light fermions given in the
previous section, we can write the scattering amplitude for
the $s$-channel pair production  $f_1 \bar f_1 \to f_2 \bar f_2$
involving fermions of definite helicity
as 
\beq
i\M =  \Big( i  \bar v_{f_1}(k_1)  \gamma^m u_{f_1}(k_2) \Big) \,
\Big( -i  \eta_{mn} S(p) \Big) \, \Big( i  \bar u_{f_2}(k_3) \gamma^n
v_{f_2}(k_4) \Big)   \ .
\eeq{eeff}
Here the subscripts $f_1$, $f_2$ denote the chiralities of the initial
and final fermions and the $s$-channel amplitude $S(p)$ is given by 
\beq
S(p) \equiv \int_{z_0}^{z_R} { dz_1 \over (kz_1)^4 } \int_{z_0}^{z_R} { dz_2 \over (kz_2)^4 } \, 
 \bigl| f_1(z_1) \bigr|^2 \, \bigl| f_2(z_2) \bigr|^2 
\Big( {\cal Q}_1 \, 
\G (z_1, z_2, p) \,  {\cal Q}_2 \Big) \ ,
\eeq{fullS}
where 
$f_{1,2} (z)$ are the zero mode 
wavefunctions of the initial and final fermions and  ${\cal Q}_1$ and
${\cal Q}_2$ are the quantum numbers of $f_1$ and $f_2$ under the
$U(1)$ symmetry.  
 (For a non-Abelian group, ${\cal Q}$ will be
generalized to a vector of gauge  charges.)  
We use the notation that ${\cal Q}$ includes the 5D gauge coupling $g_5$.

 The spinor and 
Lorentz structure of \leqn{eeff} is  the same as that in  a 4D
calculation of the same cross section.   Therefore, the effect of the
RS dynamics on each individual helicity amplitude is given simply by 
replacing the $s$-channel vector boson propagator by the quantity
$S(p)$
in \leqn{fullS}. 
In the simplest case of QED, this multiplies the scattering amplitude
by the factor
\beq
     \M / \M_{SM} =      {S(p) \over  g^2/p^2}  \ . 
\eeq{KKformfactor}
We can view this expression either as a form factor for the photon or, for $p
\ll 1/z_R$, as an unmodified photon propagator plus a contact
interaction representing the Kaluza-Klein (KK)  states. 

In this paper, we will focus on the reactions $\ee \to f \bar
f$, where the left- and right-handed 
 electrons are assumed to be approximately structureless, that is,
 associated with zero modes localized close to the UV brane. 
Then $S(p)$ simplifies to 
\beq
S(p) =  \int_{z_0}^{z_R} { dz \over (kz)^4 } 
 \bigl| f_f(z) \bigr|^2 \Big( {\cal Q}_e \, \G (z_0, z, p) \,  {\cal Q}_f \Big)  \ .
\eeq{S}
In the rest of this paper, including examples in which $\G$ is a
matrix, we abbreviate
\beq
\VEV{ \G (z, p) } \equiv \int_{z_0}^{z_R} { dz \over (kz)^4 }  \bigl|
f_f(z) \bigr|^2 \, 
            \G (z_0, z, p) \ , 
\eeq{Gmoment}
so the $S(p)$ takes the simple form
\beq
S(p) = {\cal Q}_e^T \, \VEV{\G ( z, p) }\, {\cal Q}_f \ .
\eeq{SUV}
If the fermion $f$ has a UV boundary kinetic term, 
\leqn{S} and \leqn{Gmoment} should be modified accordingly to include its effect 
as in \leqn{leftnorm} and \leqn{expformforf}. 

If the fermion $f$ is also extremely localized in the UV, $S(p)$
reduces to its form in the SM up to small corrections.   However, we
can obtain nontrivial effects if the $f$ zero mode extends into the
infrared, or, in the language of RS phenomenology, the $f$ is
{\it partially composite}.

\subsection{Pair production in RS with bulk $U(1)$ symmetry}

Using this formalism, we can compute the modification of the reactions 
$\ee \to f \bar f$ under a bulk $U(1)$ gauge symmetry. For
definiteness, we consider the case in which the $U(1)$ charges of the
zero mode fermions are equal to 1, so  that ${\cal Q} = g_5$. Then the
modification factor \leqn{KKformfactor} is the same for all four cases
of fermion helicity.   We will consider the $U(1)$ field to have a UV
boundary kinetic term with coefficient $a_B$. We will work out the
implications of this model for the various choices of $+$ and $-$
boundary conditions of
the $U(1)$ gauge field. 

First, consider the $(++)$ boundary condition. 
 The $(++)$ gauge field includes a massless photon and its KK
 resonances.   The gauge boson propagator in this case is computed in
 Appendix C of \cite{YPtwo} and found to take the  form
\beq
g_5^2 \G_{(++)} (z_1, z_2, p) =  g_5^2 k z_1 z_2 { G_{B+-}(z_<, z_0)
  G_{+-} (z_>, z_R)   \over G_{B--} } /, ,
\eeq{simpleGpp}
where $z_<$ ($z_>$) is the smaller (larger) of $z_1$ and $z_2$.
The form is rather intuitive; the Green's function satisfies the
correct boundary conditions at both the UV and the IR boundaries.  The
prescription \leqn{GBnotation} modifies the UV boundary condition on
the Green's function to 
account for the delta function kinetic term on the boundary. 
For $\ee \to f \bar f$, we set $z_1 = z_0$ and find 
\beq
S_{(++)}(p) =  {g_5^2 k \over L_B} {1  \over p^2 } \VEV{ p z \,
   { G_{+-} (z, z_R)\over G_{B--}/L_B } }\ ,
\eeq{Uoneformpp}
where we have used \leqn{LBdefin}.  The quantity in the expectation
value goes to 1 for $p \ll 1/z_R$. 

We can get further insight by expanding the expression
\leqn{Uoneformpp} for low
energy reactions. Assuming that the center of mass energy is smaller
than the scale of the IR brane, we expand the Bessel functions in
$S(p)$ around $p z_R = 0$. In this expansion, we will ignore corrections of
order $p^2 z_0^2$.  Including the first corrections in $z_R^2 = 1/k_R^2$, we obtain
 an approximate formula
\beq
S_{(++)}(p) =  g^2 \, \left[ { 1 \over p^2 } + { \delta_{KK}
    \over k_R^2 }  + \cdots \right] \ ,
\eeq{S++approx}
where we have introduced 
 the 4D gauge coupling $g^2 ={g_5^2 k / L_B }$ as in \leqn{gfourfive}.
In this low-energy effective form,  the first correction to the
massless photon exchange appears as a dimension-6 contact interaction.
The
contact interaction describes the effect of  massive KK boson exchanges. 
The strength of the contact interaction $\delta_{KK}$ is given by
\beq
\delta_{KK} = {1 \over 4} \left( - { 1 \over L_B} + \VEV{ z^2 \over z_R^2 } + 2 \VEV{{ z^2 \over z_R^2}  \log {z_R \over z}} \right) .
\eeq{contact}
Note that the effect of the gauge boundary term $a_B$ always appears
 in the form of $L_B$.  If the final state fermions are UV-localized,
 the latter two terms in \leqn{contact}  are of order $z_0^2/z_R^2$
 and $\delta_{KK}$ is 
negative. On the other hand, if the final state fermion is confined
 in the IR brane, the sum of the two terms is 1 and $\delta_{KK}$ 
 is maximized. From this, we obtain  an 
upper bound $\delta_{KK} < 1/4 $. 

It is well known that, for zero boundary kinetic terms,
 the Kaluza-Klein corrections to scattering in RS
vanish if $c_f = 1/2$.  At this value, the fermion wavefunctions in
the coordinate system \leqn{RSmetric} are constant in $x^5$ and so the
fermion currents are orthogonal to the KK gauge boson
wavefunctions~\cite{GW,DHR}.   In the presence of the boundary terms,
there are also significant cancellations.   In the $p^2$ expansion, we
find, using \leqn{expformforf},
\beq
\left( \VEV{ z^2 \over z_R^2 } + 2 \VEV{{ z^2 \over z_R^2}  \log {z_R
      \over z}} \right) \Bigg|_{c_f = 1/2} = {1\over L_\psi}
\eeqn
where $L_\psi = \log (z_R/z_0) + a_\psi$. If we further have $a_\psi =
a_B$, we find that $\delta_{KK} = 0$.   Actually, it is not difficult to
show that, 
\beq
S(p)|_{c_f = 1/2, \ a_\psi = a_B} = {g^2 \over p^2}
\eeqn
as an exact result.

Next we consider the other boundary conditions, where the $U(1)$
symmetry is broken in either UV or IR boundary. Those 5D gauge fields
do not include a massless mode. For a $(+-)$ boundary condition, 
we find
\beq
\G_{(+-)} (z_1, z_2) =  k z_1 z_2 { G_{B+-}(z_<, z_0) G_{++}(z, z_R)
\over G_{B-+}  } \ .
\eeq{Uoneformpm}
Then 
\beqa
S_{(+-)}(p) &=&  {g_5^2 k \, \over p\, G_{B-+} } \VEV{ z \, G_{++} (z, z_R) } \CR 
&=& - { g_5^2 k \, \over 2 k_R^2 } \left( 1 - \VEV{ z^2 \over z_R^2 } \right) \ .
\eeqa{S+-}
We have only the contact interaction. Notice  that the UV boundary kinetic term $a_B$ has no effect at low energy if there is no massless gauge field.  

For a $(-+)$ boundary condition, we have 
\beq
\G_{(-+)} (z_1, z_2) = k z_1 z_2 { G_{++}(z_<, z_0) \, G_{+-}(z_>, z_R) \over G_{+-} } \ .
\eeq{Uoneformmp}
Then, when the UV-localized electron is in the UV, we
have $S(p) = 0$,
 because  $G_{++}(z, z_0)$ vanishes as $z \to z_0$. 
The result is easy to understand  geometrically: A gauge field 
with $(-)$ UV boundary condition does not couple 
to a UV-localized electron. Similarly,  for the $(--)$ boundary condition, we have
\beq
\G_{(--)} (z_1, z_2) = k z_1 z_2 { G_{++}(z_<, z_0) \, G_{++}(z_>, z_R) \over G_{++} } \ .
\eeq{Uoneformmm}
and again $S(p)=0$.

\section{$SO(5) \times U(1)$ Model of Gauge-Higgs Unification}

In this section, we review the elements 
of the $SO(5) \times U(1)$ model that are essential for our study of
$\ee \to f \bar f$.
 Further details of the model can be found in \cite{YPtwo}.

\subsection{Group structure and boundary conditions}

For a realistic model, we choose its bulk gauge symmetry to be $G = SO(5) \times U(1)_X$
\cite{ACP}. Boundary conditions
 break the bulk symmetry to the SM gauge symmetry $G_{SM}
 = SU(2)_L \times U(1)_Y$ on the UV brane and to $H = SO(4) \times
 U(1)_X = SU(2)_L \times SU(2)_R \times U(1)_X$ on the IR brane. This
 model can be viewed as a dual description of an approximately
 conformal dynamics between the scales $k_R=1/z_R$ and $k = 1/z_0$ in
 four dimensions. In the dual 4D interpretation, the system
 has a global symmetry $G$, 
of which subgroup $G_{SM}$ is gauged to a local symmetry. The strongly
interacting theory spontaneously breaks $G$ to the subgroup $H$ at the
scale $k_R$. The extra $SU(2)$ factor in $H$ is a custodial symmetry
that protects the relation $m_W =  c_w m_Z$ from receiving large 
corrections~\cite{custodial}.  The study 
of the 5D model gives a calculable approach to the 4D theory.

We  label the  10 generators of $SO(5)$
as $T^{aL}$, $T^{aR}$, $T^{a5}$, $T^{45}$, with $a = 1,2,3$.   The
generators $T^{aL}$, $T^{aR}$ generate the $SO(4) = SU(2)\times SU(2)$
subgroup of $H$.  The first $SU(2)$ here is identified with the
$SU(2)$ weak interaction gauge group.   The generator of the $U(1)$ is
labelled $T^X$.  We assign the boundary condtions
\beqa 
A^{aL}_{m} &\sim & \pmatrix{+&+} \CR
A^{bR}_m &\sim &  \pmatrix{-&+}   \CR
A^{a5}_m, A^{45}_m &\sim & \pmatrix{-&-}   \ , 
\eeqa{gaugebc}
for $a = 1,2,3$, $b = 1,2$. Let $g_5$ and $g_X$ be the
5D gauge couplings of $SO(5)$ and $U(1)$.  
Introduce an angle $\beta$ such that 
\beq
c_\beta \equiv \cos \beta = {g_5 \over \sqrt{g_5^2 + g_X^2}}, \qquad
s_\beta \equiv \sin \beta = {g_X \over \sqrt{g_5^2 + g_X^2}} \ .
\eeqn
We assign the combinations
\beq  \pmatrix{ Z'_m \cr B_m } =
\pmatrix{ c_\beta & -s_\beta \cr s_\beta & c_\beta} \pmatrix{ A^{3R}_m
  \cr A^X_m } 
\eeq{BZ}  
to have the boundary conditions
\beqa
B_m &\sim&  \pmatrix{+ & +} \CR
Z'_m &\sim&  \pmatrix{-& +}   \   .
\eeqa{BZbc}
In all, there are four $(++)$ 5D gauge bosons, giving four zero modes
that can be associated with the four 4D gauge bosons of  $SU(2)\times
U(1)$. 

The boundary conditions on the $A^A_5$ gauge fields are the opposite
of those written above for $A^A_M$.  Then the fields $A^{c5}_5$ have
zero modes that are associated with scalars in 4D.   These are the
Goldstone bosons resulting from the spontaneous breaking of $G$ to
$H$. 

In terms of the gauge fields with definite boundary 
conditions, the 5D covariant derivative is
\beq
D_M = \partial_M -i\left[ g_5 A_M^{aL} T^{aL} + g_{5Y}  B_M Y
 + g_5 A_M^{bR} T^{bR} + {g_5 \over c_\beta} Z'_M (T^{3R} 
- s^2_\beta Y)  + g_5 A_M^{c5} T^{c5}  \right] \, .
\eeq{DM}
summed over $a = 1,2,3$, $b = 1,2$, $c = 1,2,3,4$. 
The 5D hypercharge coupling is given by $g_{5Y} = g_5 s_\beta$.
The hypercharge and electric charge are given by
\beq
Y = T^3_R + X \qquad \textrm{and} \qquad Q = T^3_L + T^3_R + X
\eeq{charges}
where $X$ is the $U(1)_X$ charge. 

Lastly, we include UV boundary kinetic terms $a_W$ and $a_B$ for the 
$SU(2)_L \times U(1)_Y$ bosons. From the point of view of duality, the
values of the $SU(2)_L \times U(1)_Y$ gauge couplings would be set at
some much larger energy 
scale, perhaps at the scale of grand unification. These settings would 
appear in the RS model as
boundary conditions on the UV brane. We use the boundary 
kinetic terms to parametrize the effect. 

Explicit formulae for the representation matrices described in this
section are given in Appendix B of \cite{YPtwo}.

\subsection{Identification of the Higgs field}
 
The four zero-modes $A^{a5}_5, A^{45}_5$ transform as a doublet under
$SU(2)_L$. These have precisely the quantum numbers of the complex
doublet of Higgs fields.

 Because
 the Higgs fields appear as components of gauge fields, we can gauge away
their vacuum expectation values in the central region of $z$. However,
 in a 5D system with boundaries, we cannot gauge away these
 background fields completely.  Instead, such a gauge transformation leaves
singular fields at $z_0$ or $z_R$. We can
 parametrize the gauge-invariant information of the background fields
 in terms of a Wilson line element from $z_0$ to $z_R$.  The 
Coleman-Weinberg potential of the Higgs field will depend
 on this variable~\cite{YPone}.

We can  align the expectation value along the $A^{45}$
direction.  Then the Wilson line element becomes
\beq
     U_W = \exp \left(  -i g_5 \int_{z_0}^{z_R} dz \, N_h z
   \VEV{h} 
       T^{45} \right) 
= \exp \left(- \sqrt{2} i { \VEV{h} \over f} T^{45} \right)  \  ,
\eeq{UW}
where $N_h$ is the Higgs field normalization constant:  
\beq
N_h = [(z_R^2  - z_0^2)/2k]^{-1/2}  \   . 
\eeq{normofAfive}
The Goldstone boson decay constant $f$ is analogous to the pion decay
constant in QCD. The separation between $\VEV{h}$ and $f$ represents
the `little hierarchy' between the weak interaction scale and the
scale of 5D physics.   Since the KK excitations are not yet observed, 
$\theta = \VEV{h}/f$ must be small.   Therefore, we can use 
$s = \sin \theta$ as an expansion parameter in our study of $\ee \to f \bar f$. 

\subsection{Gauge choice}

When we gauge away the Higgs field in the interior of the RS space, we
have a choice whether to move it to the UV or the IR boundary.  We
will refer to the first case, in which $U_W$ acts on the UV boundary,
as the UV gauge and to the second case as the IR gauge. 

In the UV gauge, we have definite boundary conditions for the fields
on the IR brane.  The Green's functions in this gauge are most simply
written in terms of $G$ functions pivoted
at $z_R$, \eg, $G_{\alpha \beta}(z,z_R)$, which satisfy the IR
boundary conditions manifestly.  Since these functions are already 
independent of $z_0$, this choice makes it easier to take the limit
$z_0^2/z_R^2 \to 0$, which will simplify our expressions. 

On the other hand, the IR gauge also gives attractive 
simplifications.   We have seen examples of these in the discussion of
precision electroweak corrections in \cite{YPtwo}.   It is a general
property that boundary mixing terms for two fields with identical
boundary conditions have no effect.   For example, we can apply this
to the vector fields $A^{3R}$ and $A^{X}$. These have the same
boundary condition in the IR, and so some expressions generated by the 
mixing
\leqn{BZ} disappear in the IR gauge.   Similarly, as we will discuss
below, the expressions for zero-mode wavefunctions are generally
simpler in the IR  gauge. 

It would be best if we could utilize the advantages of the
both gauge choices, using $G_{\alpha \beta}(z,z_R)$ for gauge Green's
functions as well as having unmixed boundary conditions for the
gauge field and fermion zero-modes.  A convenient prescription is to
compute Green's functions in the UV gauge but then transform to the IR
gauge using the relation
\beq
\G_{IR} (z_1,z_2) = U_W \, \G_{UV}(z_1,z_2) \, U_W^\dagger \ . 
\eeq{IRUVrelation}
Some consequences of this formula are discussed in \cite{YPtwo}, and
an explicit proof of the formula is given in Appendix D of that
paper. 

\subsection{Parametrization of the model space}

With our set-up of the $SO(5) \times U(1)$ model, the masses of the W
and Z bosons are given,
to leading order in $s^2$,  by \cite{YPtwo}
\beqa
m_W^2 &=&  { s^2 \over L_W}{1\over   (z_R^2 - z_0^2) }
 = {g_5^2 k \over 4 L_W} f^2  \sin^2 {\VEV{h} \over f } \ , \CR
m_Z^2 &=&  {L_B + s_\beta^2 L_W \over L_B } m_W^2  \ .
\eeqa{gaugemasses}
where $L_W = \log (z_R/z_0) + a_W$ and similarly for $L_B$.   For $s^2
= 0.1$ and $k_R = 1.5$~TeV, we need $L_W \approx 30$ to obtain the
measured value of the $W$ mass.   We define
basic values of 
the 4D $SU(2)$ gauge coupling $g$, the SM Higgs vacuum expectation 
value $v$, and the weak mixing  angle $\theta_w$ by
\beq
g^2 = {g_5^2k \over L_W}, \quad v = f \sin { \VEV{h} \over f}, 
\quad \cos^2 \theta_w = {L_B \over L_B + s_\beta^2 L_W} \ .
\eeq{4Dcouplings}
With these choices, the vector boson masses and cross sections obey
the tree-level SM relations.   Corrections to those relations appear
at the next order in $s^2$.   The most important of these are computed
in Section 6 of \cite{YPtwo}. 

The mass of the W boson is determined by $s^2$, $L_W$, and $z_R$. We
choose $s^2$ and $k_R = 1/z_R$ as the two main parameters of our study. $k_R$
gives the scale of the new strong dynamics and $s^2 = v^2/f^2$
parametrizes the `little hierarchy', that is, the degree of fine
tuning between the electroweak scale and the new dynamics. The $SU(2)$
boundary kinetic term 
$L_W$ allows us the freedom to fit the W boson mass for our 
given choice of $k_R$ and $s^2$. 

In an explicit RS model, the 
 value of $s^2$ is determined by the minimization of the 
Coleman-Weinberg potential for the Higgs field. In
\cite{YPtwo}, we studied a particular model of $SO(5) \times U(1)$
gauge symmetry where the top quark competes with a vector-like fermion
multiplet to generate the correct Higgs potential for the observed
Higgs mass. This strategy is general and can be applied to other
models of gauge-Higgs unification.
 Even restricting to the choice of  $SO(5) \times U(1)$
symmetry, we can have many different realistic models depending on the
field content in the RS bulk.

However, when we compute the cross sections for  $\ee \to f \bar f$ with SM final
states, the expressions that we obtain depend on these choices only
through the parameters  $s^2$ and $k_R$. Our expressions will depend on the
$SO(5)$ representations chosen for the SM  fermions, and
on the $c$ parameters of these $SO(5)$ multiplets.   But, beyond this,
our results  will be correct in full generality  for any
gauge-Higgs unification model 
based on $SO(5) \times U(1)$ symmetry.

\section{Pair-production of fermions in the $SO(5) \times U(1)$ model}

Using an expansion in the small parameters $s^2$ and $(p/k_R)^2$, 
we can obtain useful insight into the structure 
of the Green's function in the $SO(5) \times U(1)$
model. In this section, we present a general expression 
for the propagator of neutral gauge fields and 
apply it to the process $\ee \to f \bar f$ with the
simplest final states,  UV-localized massless fermions. 

\subsection{Neutral boson propagator}

Using the formalism described in \cite{YPtwo}, we can compute the
matrix $\G(z_1,z_2;p)$ in the IR gauge for
the neutral gauge fields of the $SO(5) \times U(1)$ model. The full
expression for $\G(z_1,z_2;p)$ is given in Appendix B.  
Its low-energy effective expression for the $s$-channel boson exchange is given by
\beqa
S(p)  &=& {e^2 \over p^2} \, Q_e Q_f
 + {g_{eff}^2 \over c_w^2}{ 1 \over p^2-m_Z^2} \left( T^{3L}_e - s_*^2
   Q_e \right) \left( \left(1+\delta_Z^f \right) \left( T^{3L}_f -
     s_*^2 Q_f \right)  + \delta Q_f \right) \CR
& & + {g^2 \over k_R^2} \Bigg[\delta_{KK}^W \, T^{3L}_e 
\, T^{3L}_f 
	+ { s_w^2 \over c_w^2} \, \delta_{KK}^B \, Y_e \, Y_f \Bigg] \ .
\eeqa{approxS}
The expression \leqn{approxS}  is composed of the three contributions,
associated respectively with the massless 
photon, the Z boson, and the contact interaction. The photon
propagator 
is protected from corrections. On the Z pole, the couplings to the
initial and final state fermions factorize. The couplings are
expressed in terms of $g_{eff} = g ( 1 + \delta g)$, $s_*^2 = s_w^2 (1
+ \delta s_w^2)$ and $\delta_Z$, where $g$ and $s_w^2$ are the SM
$SU(2)$ coupling and the Weinberg mixing angle. An additional `charge'
$\delta Q$ is also defined for composite final state fermions. To
order $s^2$, the 
expressions for  $\delta g$, $\delta s_w^2$, $\delta_Z$ and $\delta
Q$ are 
\beqa
\delta g &=& {m_Z^2 z_R^2 \over 4}
	 \left( {3 \over 4} - {c_w^2 \over L_W} - {s_w^2 \over L_B} \right) \CR
\delta s_w^2 &=& {c_w^2 m_Z^2 z_R^2 \over 4}
	 \left( {1 \over L_W} - {1 \over L_B} \right) \CR
\delta_Z &=& {m_Z^2 z_R^2 \over 4} \left( \VEV{ z^2 \over z_R^2 } + 2 \VEV{{ z^2 \over z_R^2}  \log {z_R \over z} } \right) \CR
\delta Q &=& \left( {s^2 \over 2} \left(-T^{3L} + T^{3R} \right) 
 	+ {s \over \sqrt{2}} T^{35} \right) \VEV{ z^2 \over z_R^2 } \ .
\eeqa{devs}
It is instructive to note that $\delta Q$ is the only term that 
includes an explicit dependence on $T^{3R}$ and $T^{35}$. All
other terms of $S(p)$ are written in terms of the $SU(2) \times U(1)$
quantum numbers $T^{3L}$ and $Y$. We will see below that the quantum number 
$T^{35}$ does not contribute to light fermion matrix elements.

The contact interactions are parametrized by $\delta_{KK}^{W}$ and
$\delta_{KK}^{B}$, which 
are defined analogously to \leqn{contact},
\beqa
\delta_{KK}^W &=& {1 \over 4} \left( - { 1 \over L_W} + \VEV{ z^2 \over
    z_R^2 } + 2 \VEV{{ z^2 \over z_R^2} 
 \log {z_R \over z}} \right) \ , \CR
\delta_{KK}^B &=& {1 \over 4} \left( - { 1 \over L_B} + \VEV{ z^2 \over
    z_R^2 } + 2 \VEV{{ z^2 \over z_R^2} 
 \log {z_R \over z}} \right) \ .
\eeqa{contactW}
They originate from the KK states
of the two $(++)$ bosons, $A^{3L}$ and $B$. As the UV-localized
initial electron does not couple to the gauge fields with $(-)$ UV
boundary conditions, the KK states of $Z'$ and $A^{35}$ make no contributions. 

It is instructive to compute the expectation values for fermion zero-modes with limiting values of the $c$ parameter.
\beq
\matrix{ & c \ll -1/2 & c = -1/2 & c = 1/2 & c \gg 1/2 \cr 
\VEV{ z^2 \over z_R^2 } & 1 & 1/2 & {1 / (2L_t)} & 0 \cr
2 \VEV{{ z^2 \over z_R^2} \log {z_R \over z}} & 0 & 1/4 & 1/(2L_t) & 0 } .
\eeq{climits}
Note that $L_t$ evaluated at $c=1/2$ is given by $L_t = \log(z_R/z_0) + a_t$.

In~\cite{YPtwo}, we have seen that a limit on the oblique parameter $S
< 0.135$ gives the constraint $1/z_R > 1.5$~TeV. For $1/z_R = 1.5$ TeV,
${m_Z^2 z_R^2 / 4} = 0.09\%$. Then  deviations from $\delta g$, 
$ \delta s_w^2$, and $\delta_Z$ have only secondary importance and neglecting
them  would not change the qualitative behavior of the cross
section deviation of our model compared to the SM.   However,
significant changes can be generated by the $\delta_{KK}$ and $\delta
  Q$ terms. 
In our discussion of the cross section deviations below, we will
display
 the effects of those two terms for various assignments of the
 fermions.  We will see that the 
$\delta Q$ contribution is dominant near the $Z$ pole,  while 
the contact interaction becomes more apparent as we increase
 the center-of-mass energy.
 
\begin{figure}
\centering
\includegraphics[width=4in]{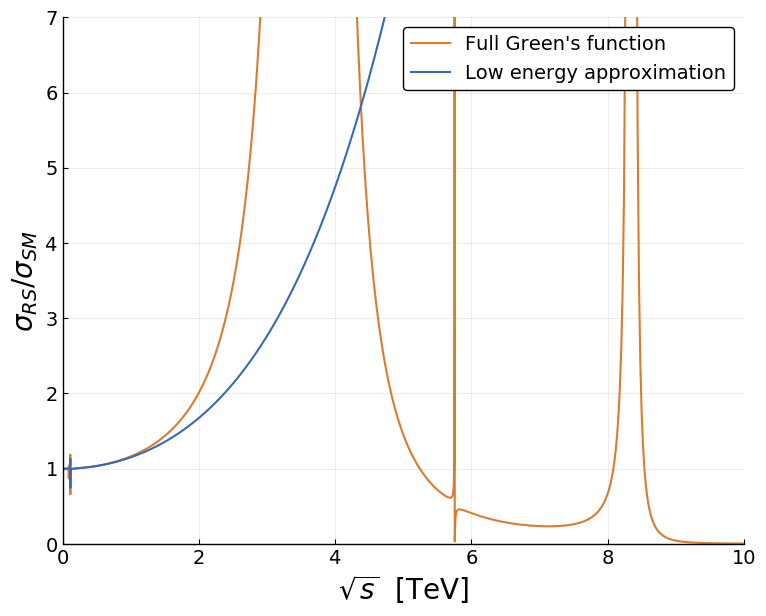}
\caption{Cross section deviations of $e^-_L e^+_R \to b_R {\bar b}_L$
  where $b_R$ is embedded in {\bf 5} of $SO(5)$. More details of the
  bottom quark embedding are explained in Section 5. Parameters used
  in this plot are $c_b = 0.3$, $s^2 = 0.1$, and $k_R = 1.5$~TeV.
The small `jitter' near $\sqrt{s} = 0.1$~TeV corresponds to the $Z$ pole.}
\label{fig:example}
\end{figure}

If we raise the center-of-mass energy further, the behavior of our
approximate formula \leqn{approxS} will eventually deviate from that
of the full Green's function \leqn{decomS}, which includes resonances
from gauge KK states at masses of a few $k_R$. Fig.~\ref{fig:example}
shows an example of how the cross section predictions differ between
the approximate formula (blue line) and the full Green's function
(orange line) for $e^-_L e^+_R \to b_R {\bar b}_L$ processes. The
vertical axis shows the size of the cross section compared to that of
the SM prediction and we used $k_R = 1.5$~TeV for this plot. We can
see that both lines start at $\sigma_{RS} / \sigma_{SM} = 1$ but
deviate near $\sqrt{s} \gsim k_R$. The KK resonances appear near 4, 6,
and 8 TeV in the full Green's function. The good agreement between the
two lines show that our formula \leqn{approxS} is indeed an excellent 
 approximation in the region of our interest, 250~GeV~$\le \sqrt{s}
 \le$~1~TeV.
 We will study the process $\ee \to b {\bar b}$ more in detail in Section 5.

\subsection{Pair-production of UV-localized fermions}

The simplest case to which we might apply the formulae in the previous
section is that of light fermions whose zero-mode wavefunctions are
strongly localized near the UV boundary.

For a UV-localized final state, $\delta Q = 0$ as implied by
\leqn{climits}. Then $T^{3L}$ and $Y$ are the only quantum numbers
relevant in the process $\ee \to f \bar f$. That is, we find the same
cross section regardless of how we embed those light flavors into an
$SO(5)$  multiplet.

For the contact interactions, we find
\beq
\delta_{KK}^W = -{ 1 \over 4L_W} \, , \, \qquad \delta_{KK}^B = -{ 1 \over 4L_B} \ .
\eeqn
which are suppressed by $1/4L_{W,B}\sim 0.01$. The 
cross section deviations from these terms and the other terms 
with ${m_Z^2 z_R^2/ 4}$ are small enough to satisfy any current
 precision constraints on the light flavors.   For $k_R > 1.5$ TeV,
 our formulae predict KK recurrences of the photon and $Z$ with masses
 above 3.5~TeV.  These resonances also have suppressed couplings to
 UV-localized zero mode fermions.

The most direct bound on $k_R$  from the LHC is that  
from constraints on  contact
interactions. The strongest lower bound now quoted for a $\Lambda$
scale in contact interactions is 40~TeV (25~TeV for destructive
interference), for a universal left-handed $\ell\ell q q$ interaction,
from the ATLAS experiment at 13~TeV~\cite{ATLASLambda}.
Translating the
limit $\Lambda > 40$ TeV into 
that of $k_R$ using  \leqn{approxS}, we obtain  $k_R  >  1$ TeV.
This is not yet as strong as the constraint from precision electroweak
measurments.

\section{Mass generation and pair-production for the $b$ quark}

In this section, we study schemes of mass generation for the bottom
quark and their  implications for  the cross section for $\ee \to b \bar
b$. The strong top Yukawa coupling and the large mass separation
between the top and bottom quark restrict how the bottom quark is
embedded into an $SO(5)$ multiplet. 

We assume that the generation of the  top quark mass 
is the main driving force of the
electroweak symmetry breaking. To achieve this, the Higgs field must 
couple with full strength to the top quark.  This implies that the 5D
multiplet containing the  $(t_L,b_L)$ doublet must also contain the
$t_R$, so that the two chirality states of the top quark are directly
linked by the 
the Higgs field $A^{45}_5$. Furthermore, the $c$ parameter of the top quark multiplet
should not be much bigger than $1/2$ so 
that it can couple strongly to the Higgs field. 
We consider the range $0.3 < c_t < 0.6$ in our analysis. 

If  $b_R$ were also  included in the top quark
multiplet, it would gain the same mass from electroweak symmetry
breaking as 
the top quark. To otain a much smaller mass for the $b$ quark, we need
to place the $b_R$ in a different multiplet and  postulate a flavor
mixing between this 5D multiplet and the top quark multiplet that
assists the Higgs field to connect the $b_L$ and $b_R$. 
We model this mixing by an $SU(2) \times U(1)$-invariant
mixing of the two multiplets on the UV brane, parametrized by $s_b = \sin
\theta_b$.

It follows from this setup that the $SO(5)$ assignment of the $b_L$ is
directly given by that of the top quark, and that the assignment of
$b_R$ is specified almost uniquely once we choose the $SO(5)$
representation of its 5D multiplet. In this paper, 
we consider two choices, $\bf 5$ and $\bf 4$ of $SO(5)$.

\subsection{$t,b$ in the {\bf 5} of $SO(5)$}

In \cite{YPtwo}, we assigned the mutiplet $\Psi_t$ that contains the
$(t_L, b_L)$ multiplet to the {\bf 5} of $SO(5)$.  This follows the 
suggestion by Agashe, Contino, Da Rold, and
Pomerol that this choice  provides a custodial symmetry constraining the $Zb\bar b$
coupling~\cite{Zbb,5rep}.   We found that this choice also has
quantitative advantages in fitting the mass of the Higgs boson~\cite{YPtwo}.   We
can also embed the $b_R$ into a different {\bf 5} multiplet
$\Psi_b$.  
 
More specifically, we embed the 
top and bottom quarks as
\beq 
\Psi_t = \left[ \matrix { \pmatrix { \chi_f (-+) & t_1 (\Box +) \cr 
	\chi_t (-+) & b_1 (\Box +)} \cr t_R (--)} \right]_{X=2/3}, \quad 
\Psi_b = \left[ \matrix { \pmatrix { t_2 (\Box +) & \chi_b (-+) \cr 
	b_2 (\Box +) & \chi_f'(-+)} \cr b_R (--)} \right]_{X=-1/3} \ .
\eeq{t5}
We will denote the 5D mass parameters
 of the two multiplets as $c_t$ and $c_b$, respectively. In \leqn{t5},
 the matrix in parentheses is a bidoublet, with $SU(2)_L$ acting
vertically and $SU(2)_R$ acting horizontally.  The fields $\chi_f$ and
$\chi'_f$ have electric charge $Q=5/3$ and $Q=-4/3$,
respectively. The signs in parentheses indicate the UV and IR boundary
conditions for each field.  The $t_R$ and $b_R$ fields contain
 right-handed zero modes. The notation $\Box$ denotes mixing on the
UV brane, as we will now explain.

For the left-handed zero mode $t$ and $b$ fields, we define the combinations
\beq  \left[ \matrix{ (t_L, b_L) \cr (t'_L, b'_L) } \right] =
\pmatrix{ \cos \theta_b & -\sin \theta_b \cr \sin \theta_b & \cos \theta_b} 
\left[ \matrix{(t_1, b_1) \cr (t_2, b_2)  } \right]
\eeq{tbmixing}  
and assign definite boundary conditions 
\beq
\matrix { t_L (++) \cr b_L (++) }   \qquad 
\matrix { t'_L (-+) \cr b'_L (-+) } \ .
\eeq{tbbc}
Note that $(t_1,b_1)$ and $(t_2,b_2)$
have the same $T^{3L}$ and the same $Y = T^{3R}+X$, so
 the mixing is $SU(2)_L \times U(1)_Y$ invariant. 
The $(t_L, b_L)$ contain  left-handed zero modes. Note that $(t_L, b_L)$
and $(t'_L, b'_L)$ must be assigned opposite UV boundary conditions;
otherwise the mixing has no effect.   This construction allows the
Higgs field, which acts only within an $SO(5)$ multiplet, to connect
$b_L$ and $b_R$, generating a $b$ quark mass. 
 
We also include the UV boundary kinetic term $a_t$ for the $(t_L,
b_L)$ doublet, so that we can adjust this parameter to obtain the 
 correct top quark mass for any values of $c_t$. The details of the 
formalism of the fermion UV boundary kinetic term can be found in
\cite{YPtwo}. 

In the limit where the $\theta_b$ effect is a small perturbation, the
mass of the top 
quark is determined by the 
four model parameters: $s^2 = v^2/f^2$, $z_R$, $c_t$, and $a_t$.  The mass
of the $b$ quark depends on the two additional parameters $c_b$, and
$\theta_b$. 
The full formulae
 for the $t$ and $b$ masses are given in Appendix C. We will quote
 here the simplifications of these formula
using $m_b/m_t \ll 1$ and $z_0/z_R \ll 1$. 

 The top quark mass is 
 given  by 
\beq
    m_t^2 =   {2c_t+1\over 2}  { s^2 z_R^{-2} \over L_t}
    \biggl({z_0\over z_R}\biggr)^{c_t - 1/2} \ ,
\eeq{topmass}
with 
\beq 
   L_t =     {1\over 2c_t - 1}\biggl[  \bigl({z_R\over z_0}\bigr)^{c_t - 1/2}
   -  \bigl({z_0\over z_R}\bigr)^{c_t - 1/2}\biggr]  
+ a_t \bigl({z_R\over z_0}\bigr)^{c_t - 1/2}  \  . 
\eeq{Ltval}
For $c_t = 0.5$, $L_t = \log (z_R/z_0) + a_t$, similar to $L_B$ in
\leqn{gfourfive}.  For our choice of parameters, the logarithm equals 4.61, and
we need  values of $L_t$ in the range  5--9 to fit the observed top
quark mass~\cite{YPtwo}.  A large value of $L_t$  pulls a large fraction of 
 the $t_L$ and $b_L$
wavefunctions to the UV brane and so decreases the observable RS
effects on cross sections.  Keeping the top quark mass fixed as $s^2$,
$z_R$ and $c_t$ are varied requires a compensatory variation of $a_t$,
and this affects the predictions of the theory, as will be apparent
below.   The boundary kinetic term must have a positive coefficient,
so we will restrict ourselves to the region where $a_t > 0$ for our phenomenological
discussion. 

Using the same approximations, the 
 mass ratio between the top and bottom quark is given by 
\beq
{m_b^2 \over m_t^2} = \tan^2 \theta_b \left( {1+2c_b \over 1+2c_t} \right) \left( z_0 \over z_R \right)^{2c_b - 2c_t} \ .
\eeq{mbmt}
The relation \leqn{mbmt} 
 shows us that there are two independent strategies to realize the
correct $m_b/m_t$ ratio.  First, we can choose
 a small UV mixing angle $\theta_b$. Second, we can choose $c_b- c_t >0 $,
 giving a suppression that is exponential in this difference.   The
 first mechanism is rather intuitive, but the second is not.  It works
 because the $b_2$ field is pushed to the UV, minimizing its overlap
 with the $b_R$.  This same parameter choice pushes the right-handed
 zero mode in the $b_R$ to IR, increasing its degree of
 compositeness.

In computing the effects of the mass generation on the $\ee\to b \bar
b$ cross sections, we find that the explicit effects of $\theta_b$ are
of order $(m_b z_R)^2$, parametrically smaller than the effects
described above.   Once we have understood the ranges of $c_t$, $c_b$
expected from the mass generation mechanism, we can ignore explicit
effects of $m_b$ and $\theta_b$ in the calculation of cross sections.
This also allows us to ignore the $T^{35}$ term in \leqn{devs}. 

With this choice of assignments, $T^{3L} = T^{3R}$ for both $b_L$ and
$b_R$.   Then, 
 the parameter $\delta Q$ in \leqn{devs} equals zero up to ignorable
 matrix elements of $T^{35}$.   We will see the implication of this
 for the $\ee\to b\bar b$ cross sections in Section~5.3.

\subsection{$t,b$ in the {\bf 4} of $SO(5)$}

Another possibility is to embed the top and bottom quarks 
in the {\bf 4} of $SO(5)$~\cite{4rep},
\beq 
\Psi_t = \left[ \matrix { t_1 (\Box +) \cr
b_1 (\Box +) \cr t_R (--) \cr b'(-+)} \right]_{X=1/6}, \qquad 
\Psi_b = \left[ \matrix { t_2 (\Box +) \cr
 b_2 (\Box +) \cr t' (-+) \cr b_R (--)} \right]_{X=1/6} \ .
\eeq{t4}
The notation for boundary conditions is the same here as in \leqn{t5},
and again we represent the $t_L$ and $b_L$ mixing by \leqn{tbmixing}
and \leqn{tbbc}. The formulae for the top and bottom quark masses in
this case are also worked out in  Appendix C. The final result for the
mass ratio $m_b/m_t$ is again \leqn{mbmt}. So, also in this case,  we 
need small $\theta_b$ or $c_b > c_t$ to obtain 
 the correct mass ratio between the top and bottom quarks.

With this embedding,  we now have nonzero $\delta Q$:
\beq
\delta Q^{\bf 4}_{b_L} = {s^2 \over 4} \VEV{ z^2 \over z_R^2 }, \quad  
\delta Q^{\bf 4}_{b_R} = -{s^2 \over 4} \VEV{ z^2 \over z_R^2 } \ .
\eeq{dQ4}
We can relate $\delta Q^{\bf 4}_{b_L}$ to the top quark 
mass since $b_L$ resides in the same multiplet with $t_L$ and $t_R$. 
As shown in Appendix C, the relation is 
\beq
\delta Q^{\bf 4}_{b_L} = { m_t^2 z_R^2 \over (1+2c_t) (3-2c_t) }
\eeq{dQ4L}
for small $s^2$ and $0.3 < c_t < 0.6$. As a representative value, 
$\delta Q^{\bf 4}_{b_L} = 0.0033$ for $k_R = 1.5$~TeV and $c_t =
0.5$. 
Note that its dependence on $c_t$ is weak. Although the value of 
$\delta Q^{\bf 4}_{b_L}$ is small, the presence of nonzero $Q_{b_L}$
 is strongly constrained by $Z$ 
pole precision measurements.  We will show now that $b_L$ 
embedded in {\bf 4} of $SO(5)$ is unfavored.

In \cite{YPtwo}, we studied the constraints on the RS wavefunctions of
the $b_L$ and $b_R$ in the {\bf 5} 
arising from the values at the $Z$ resonance  of 
$ R_b$, the fraction of hadronic $Z$ decays to $b\bar b$, and $A_b$,
the polarization asymmetry in $Z\to b\bar b$.   The value of $R_b$ is
very accurately known \cite{LEPEWWG},
\beq
    R_b  = 0.216\pm 0.00066 \   .
\eeq{Rbval}
 The correction to
$R_b$ is given by 
\beq
        \Delta R_b \approx 2  R_b (1- R_b) \left(  \delta g_{Zb_L} + 
        { g_{Zb_R}^2 \over g_{Zb_L}^2}  { \delta g_{Zb_R} } \right) \ ,
\eeqn
where $\delta g_{Zb_{L,R}}$ are fractional deviations of $g_{Zb_{L,R}}$. 
This is dominated by the correction to the $b_L$ 
coupling because of the small size of the SM $Z$ coupling to $b_R$, 
\beq
        {   g_{Zb_R}^2 \over g_{Zb_L}^2}\biggr|_{SM}  =   0.033  \ . 
\eeqn
The constraints on $g_{Zb_L}$ and $g_{Zb_R}$ from $A_b$ are much
weaker. 

In RS models of the type we are describing, there 
 are three contributions to  $\delta g_{Zb_{L,R}}$, arising,
 respectively,
from  $\delta s_w^2$, $\delta_Z$
and $\delta Q$ in \leqn{devs}.   In \cite{YPtwo}, we showed that
contributions from the first two sources lead to effects that are much
smaller than the experimental error in  \leqn{Rbval}.    For $b_L$ in
the {\bf 5}, this is the end of the story.  For $b_L$ in the {\bf 4},
$\delta Q_{b_L}$ is nonzero and \leqn{Rbval} leads to the bound
\beq
|\delta Q_{b_L}| < 0.00165 \quad \textrm{(95\% CL)} \  .
\eeq{QbLbound}
which excludes the value of $Q^{\bf 4}_{b_L}$ found above.   The
effect of $\delta g_{Zb_R}$ strengthens this bound.   Including this
effect, values $k_R > 3$~TeV are needed to be consistent with our
precision knowledge of $R_b$.

However, the possibility is still open  to embed $b_L$ in the {\bf 5} but
$b_R$ in the {\bf 4}. We can mix the $(t_1, b_1)$ of $\Psi_t$ in \leqn{t5}
with the $(t_2, b_2)$ of $\Psi_b$ in \leqn{t4}, and generate the
bottom quark mass.  In this case, the formula for the $b/t$ mass ratio
becomes
\beq
{m_b^2 \over m_t^2} = \half \tan^2 \theta_b
 \left( {1+2c_b \over 1+2c_t} \right) \left( z_0 \over z_R \right)^{2c_b - 2c_t} \ .
\eeq{mbmt2}
This makes only a minor change in the overall logic.
 In this case, $\delta Q^{\bf 5}_{b_L}$ is
identically zero.  The 
bound on $\delta Q^{\bf 4}_{b_R}$ is weaker  than that quoted above due to the
suppression factor $g_{Zb_R}^2 / g_{Zb_L}^2$.
 With $\delta g_{Zb_L} = 0$, we have
\beq
|\delta Q_{b_R}| < 0.00904 \quad \textrm{95\% CL} \ .
\eeq{QbRlimit}
This should be compared to the formula \leqn{devs},
\beq
\delta Q^{\bf 4}_{b_R} = - {s^2 \over 4} \VEV{{z^2\over z_R^2}} \ .
\eeq{Qfor4R}
This formula  contains $s^2$ as the prefactor rather 
than $(m_t z_R)^2$ as in \leqn{dQ4L}; the  coefficient is larger by a
factor of $L_t \sim 5$. Computing $\VEV{z^2/z_R^2}$ for $b_R$, we have
\beq
\delta Q^{\bf 4}_{b_R} = - {s^2 \over 4}  \left( 1+ 2 c_b\over 3+ 2c_b \right)
\ .
\eeq{Qfor4Reval}
Then we find $s^2 < 0.081$ at  $c_b =
0.3$, $s^2 < 0.069$ at $c_b = 0.6$.   This is a  significant
constraint but one that allows most of the parameter space of interest
to us.

\subsection{Deviations of the pair-production cross section}

We can now assess the effects of the RS structure on the cross
sections for $\ee\to b\bar b$.   In the approximation in which we
ignore the $b$ quark mass, the  leading order cross sections for completely
polarized $\ee$ beams take the form
\beqa
{d \sigma\over d \cos \theta} (e^-_Le^+_R \to b {\bar b}) &=&
   \Sigma_{LL}(s)  \ (1 + \cos\theta)^2 + \Sigma_{LR}(s)
   (1-\cos\theta)^2 \CR
{d \sigma\over d \cos \theta} (e^-_Re^+_L\to b {\bar b}) &=&
   \Sigma_{RL}(s)  \ (1 - \cos\theta)^2 + \Sigma_{RR}(s)
   (1+\cos\theta)^2
\eeqa{basiccs}
The terms $\Sigma_{LL}$, $\Sigma_{RL}$ are associated with the
production of $b_L\bar b_R$; the terms $\Sigma_{LR}$, $\Sigma_{RR}$
are associated
 with the
production of $b_R\bar b_L$.  At a linear collider with polarized
beams and using vertex charge to distinguish $b$ and $\bar b$, all
four of these functions can be measured independently at a fixed
$\sqrt{s}$~\cite{Poschl}.

As we have explained in the previous section, the cross sections in
RS models will differ from those computed in the SM through the
effects parametrizd by $\delta_{KK}$ and $\delta Q$ in \leqn{devs}.
We consider first the case in which both the $b_L$ and the $b_R$ are
assigned to {\bf 5} representations of $SO(5)$.   Here  $T^{3L} =
T^{35}$ for both chiral states of the $b$ quark, and so  $\delta
Q^{\bf 5} = 0 $ for both $b_L$ and $b_R$.    The visible effects of
the RS structure then come only from the contact term induced by the
gauge KK states.

\begin{figure}
\centering
\includegraphics[width=5in]{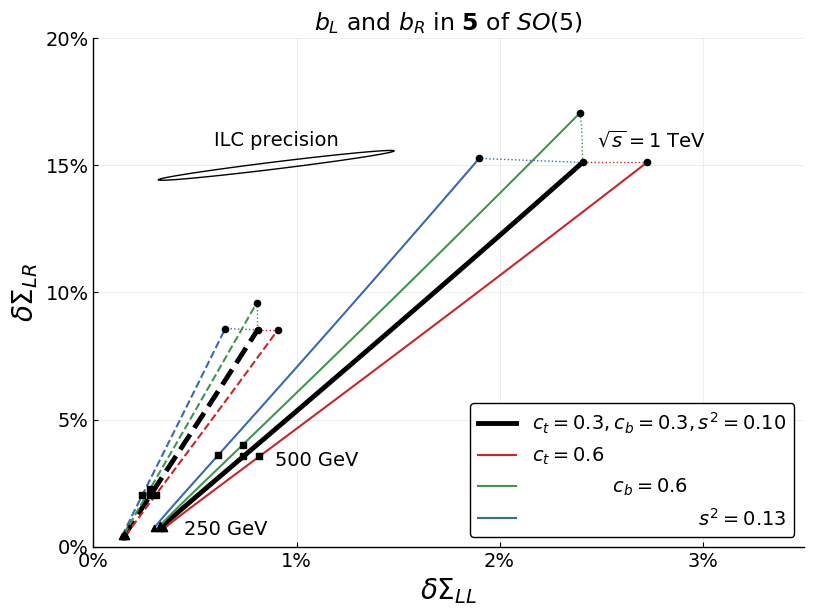}
\includegraphics[width=5in]{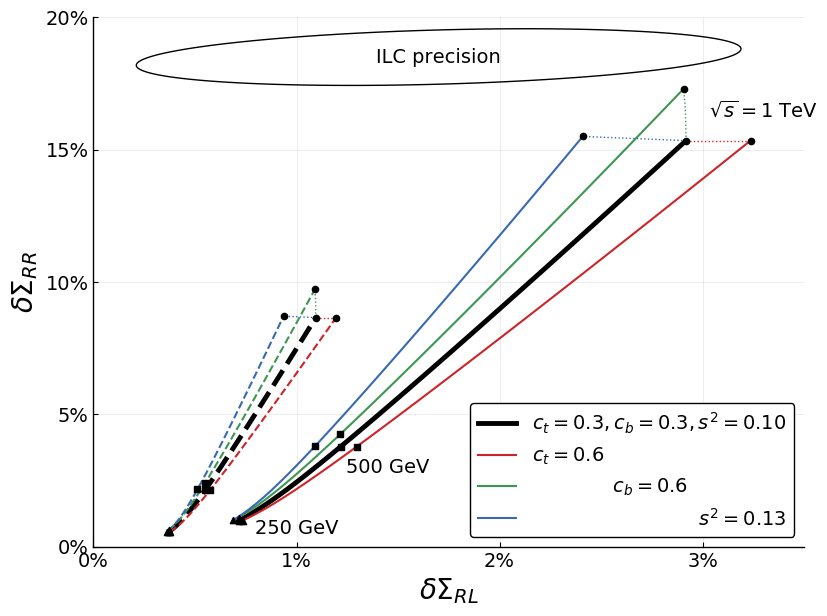}
\caption{Cross section deviations of $\ee \to b {\bar b}$. Both of $b_L$ and $b_R$ are embedded in {\bf 5} of $SO(5)$. Solid lines for $k_R = 1.5$~TeV. Dashed lines for $k_R = 2$~TeV. }
\label{fig:b5}
\end{figure}

Fig.~\ref{fig:b5} shows cross section deviations of the process $\ee
\to b {\bar b}$ for $b$ in the {\bf 5} of $SO(5)$.  The upper graph 
refers to $e^-_Le^+_R $  beams, and the lower graph to $e^-_Re^+_L $
beams.   The horizontal and vertical axes in the upper plot show the cross section
deviations 
\beq
          \delta \Sigma_{LL} = {\Delta \Sigma_{LL}\over
            \Sigma_{LL}|_{SM}}  \quad    \delta \Sigma_{LR} = {\Delta \Sigma_{LR}\over
            \Sigma_{LR}|_{SM}}  \ , 
\eeqn
and, similarly, the axes of the lower plot give the deviations in
$\Sigma_{RL}$  and $\Sigma_{RR}$.   The central black line show the
effect on these cross section contributions as the $\ee$ center of mass
energy is varied from  250~GeV to 1~TeV for the choice of parameters
\beq
      k_R = 1.5~\mbox{TeV}\ , \quad c_t = 0.3 \ , \quad c_b = 0.3 \ ,
      \quad  s^2 = 0.1  \ .
\eeq{centralparams}   
The additional parameters of the model are fixed from the precision
electroweak measurements, including the masses of the $W$ and $Z$,
and, to fix  $a_t$, the mass of the top quark.
The added lines show the effect of varying the parameters in \leqn{centralparams}
individually,  from 0.3 to 0.6 for $c_t$ and $c_b$, and from 0.1 to 0.13
for $s^2$. It should be noted that $a_t$ is always positive in this range 
of parameters. Finally, the dashed lines show deviations for $k_R = 2.0$~TeV. 
Increasing $k_R$ decreases the effects uniformly as $k_R^{-2}$. 

In this case with $\delta Q=0$, the deviations
are almost zero at small $\sqrt{s}$, but increase with $\sqrt{s}$
as the contact interaction becomes more dominant.  
With the positive value of $c_b$, the $b_R$ is
very composite and gives larger values for the moments in
$\delta_{KK}$. As an example, the value of $\VEV{z^2/z_R^2}$ 
for $c_t = c_b = 1/2$ is larger for $b_R$ than for $b_L$ 
by a factor of $L_t \sim 5-9$, as shown in \leqn{climits}. This is reflected 
in the ratio of the vertical to the horizontal scales on the plot.
 The increasing $c_b$ makes the $b_R$ more composite,
and it makes $\delta \Sigma_{LR, \, RR}$ larger.   The situation for $c_t$ is more
subtle.  One might guess that increasing $c_t$ makes the $b_L$ less
composite and therefore decreases the effect on the cross section.
However, according to \leqn{topmass}, increasing $c_t$ at fixed top
quark mass requires a decrease in the value of $a_t$.   This is
actually a larger effect in the opposite direction that makes the
$b_L$ wavefunction larger at large values of $z$.   This leads to
larger cross section deviations, as shown  in the figure. 
Similarly, increasing $s^2$ requires larger $a_t$,  and this decreases the
cross section deviation for $b_L$.

In the two  graphs in  Fig.~\ref{fig:b5}, we show our estimate of the 68\% confidence
region for measurements of the two $e^-_L$ cross sections  and the two $e^-_R$ cross sections that would be obtained at
 the International Linear Collider (ILC)  at 250~GeV with 2~ab$^{-1}$
of data.  This estimate is based on results presented in the
study \cite{Poschl}.   It will be difficult to discern these cross
section deviations at 250 GeV, though they will become apparent at
higher $\ee$ center of mass energies.

Consider next the case in which the $b_L$ is assigned to the {\bf 5}
of $SO(5)$ while the $b_R$ is assigned to the {\bf 4}.  Now 
$\delta Q^{\bf 4}_{b_R}$  is nonzero.   The effect of $\delta
Q^{\bf 4}_{b_R}$ on $\Sigma_{LR}$, $\Sigma_{RR}$ is largest on the $Z$
pole and decreases as we increase the center-of-mass
energy. This leads to a different pattern of deviations 
for the cross sections for $b_R$ production.

\begin{figure}
\centering
\includegraphics[width=5in]{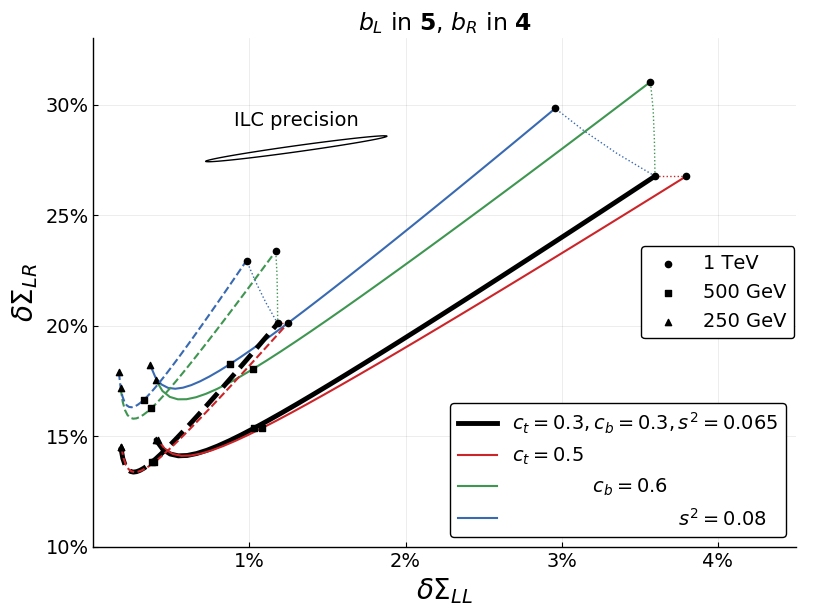}
\includegraphics[width=5in]{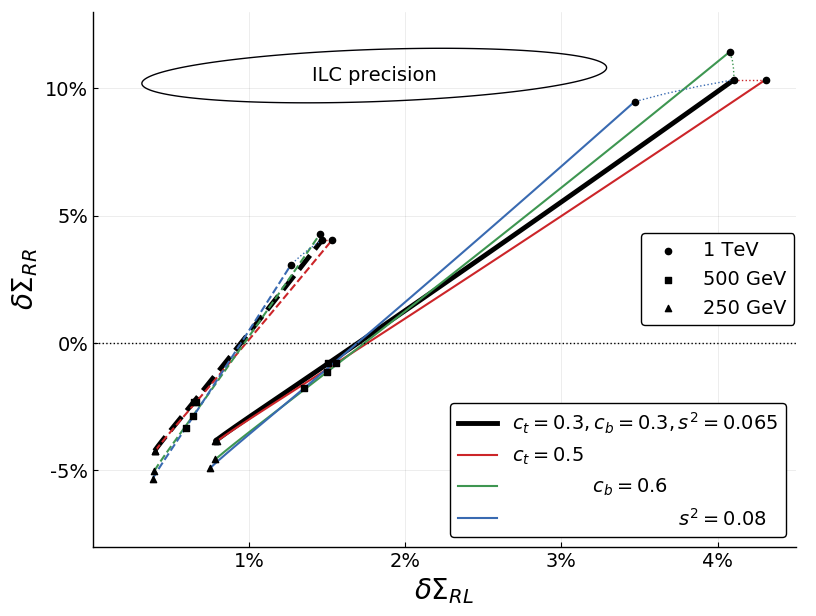}
\caption{Cross section deviations of $\ee \to b {\bar b}$, for $b_L$ in {\bf 5} and $b_R$ in {\bf 4} of $SO(5)$.  Solid lines for $k_R = 1.5$~TeV. Dashed lines for $k_R = 2$~TeV. Note that the vertical axis is plotted differently for $\delta \Sigma_{LR}$ and $\delta \Sigma_{RR}$.}
\label{fig:b4}
\end{figure}

Fig.~\ref{fig:b4} shows the cross section deviations for this case,
using the same notation as that used in Fig.~\ref{fig:b5}.   
 We see that  $\delta \Sigma_{LR}$ is  already sizable at $\sqrt{s} =
 250$~GeV, enough so to be readily observed at the 250~GeV ILC. 
As we go away from the $Z$ pole, the effect of $\delta Q^{\bf
  4}_{b_R}$ decreases and the contact term becomes more dominant. This
explains the non-monotonic behavior of $\delta \Sigma_{LR}$ as
$\sqrt{s}$ is increased. Also note that $\delta \Sigma_{LR}$ is not
suppressed with  larger $k_R$ (dashed lines) at small $\sqrt{s}$,
since $\delta Q^{\bf 4}_{b_R}$ does not depend 
on $k_R$ directly.

It is interesting to see that the sign of the the effect of $\delta Q^{\bf 4}_{b_R}$
changes between the  $e^-_L e^+_R$ and the 
$e^-_R e^+_L$ initial state.    To understand this, look back at \leqn{approxS}. 
Due to the small $Z$ coupling of
 $b_R$, the photon propagator is dominant over the $Z$ propagator for
 $b_R {\bar b}_L$ final state.   According to \leqn{devs}, 
 $\delta Q^{\bf 4}_{b_R}$ is negative.   In \leqn{approxS}, the term
 with $\delta Q$ is multiplied by $(T^{3L}_e - s_w^2 Q_e)$, which is
 negative for $e^-_L$ and positive for $e^-_R$.  Comparing to the
 photon propagator term, this contributes constructively for $e^-_L$
 and destructively for $e^-_R$, in accord with the results displayed
 in the Fig.~\ref{fig:b4}.

Although Figs.~\ref{fig:b5}  and \ref{fig:b4}  include only  slices of
the paramter
space, they  allow us a detailed understanding of how each parameter
affects the cross section.  They also demonstrate that the two cases
have  physically distinct predictions and can be distinguished by
precision experiments.    The
 most important effect appears in the backward cross section for
 $e^-_Le^+_R$ beams, so both beam  polarization and excellent $b/\bar
 b$ separation is needed to observe and separate the predicted
 effects.

\section{Pair-production of the $t$ quark}

In this section, we compute the pair production cross section for  the top
quark. In the calculations of the previous section, we made strong use
of the fact  
that the bottom  quark mass
is much smaller than the new physics scale $k_R$ and the 
center-of-mass energy $\sqrt{s}$ even at 250~GeV. This allowed us
to ignore the effect of the mass in the pair production cross
section.    However, for the massive
 top quark, we must  take the nonzero quark mass into account.   This
 does not change the calculation conceptually, but it adds more
 bookkeeping that must be carried out correctly.

\subsection{Cross section calculation for the top quark}

The calculation of the $\ee\to t\bar t$ cross section adds four new
elements.
First, the mass of the top quark cannot be ignored, and so both the
left- and right-handed chirality states of the top quark must be
included.  Second, each top quark chirality  state is a part of a 5D
field that is a Dirac fermion, and so, both for $t_L$ and $t_R$, the
full 4-component Dirac fermion must be accounted.  Third, the operator
$T^{35}$ in \leqn{DM} and \leqn{devs}, which has nonzero matrix elements
between the $t_L$ and $t_R$ Dirac fermions, must now be taken into
account.  
Finally, the $t\bar t$ final states have four possible
helicity states, so \leqn{basiccs} must now be enlarged to 
\beqa
{d \sigma\over d \cos \theta} (e^-_Le^+_R \to t {\bar t}) &=&
   \Sigma_{LL}(s)  (1 + \cos\theta)^2 + \Sigma_{LR}(s)
   (1-\cos\theta)^2 + \Sigma_{L0} (s) \sin^2\theta\CR
{d \sigma\over d \cos \theta} (e^-_Re^+_L \to t {\bar t}) &=&
   \Sigma_{RL}(s)  (1 - \cos\theta)^2 + \Sigma_{RR}(s)
   (1+\cos\theta)^2 + \Sigma_{R0} (s) \sin^2\theta \, . \CR
\eeqa{basiccsfort}
In this equation, the LL and RL terms are associated with the
production of the helicity (not chirality) state $t_L\bar t_R$,  the
RL and RR terms are associated with the
production of the helicity state $t_R\bar t_L$,  and the   L0 and R0
terms are associated with the production of the helicity states
$t_L\bar t_L$ and $t_R\bar t_R$.  In the models discussed in this
paper, the latter two helicity amplitudes are equal by $CP$. 

One more possible complication does not appear.  The 5th
components of the gauge fields have couplings of the form $\bar\Psi
\gamma^5 \Psi$ with nonzero 
 matrix elements between the left- and right-chirality
components of the 5D fermion fields.  However, as we noted already in
Section~2, our use of the Feynman-Randall-Schwartz gauge
\cite{RandallSchwartz}
implies that, since the massless electron does not couple to the
$A^{A}_5$ fields, the diagrams in which the $A^A_5$ fields couple to the
top quark are zero. 

To compute the amplitudes for  $\ee\to t\bar t$, we must first
construct the $t$ and $\bar t$ wavefunctions, and then take the matrix
elements   between these wavefunctions of the propagator \leqn{approxS}.
In the $SO(5)\times U(1)$ model, with $t_L$ and $t_R$ assigned to the
{\bf 5} of $SO(5)$, the $t_L$ and $t_R$ belong to the
multiplet
$\Psi_t$  in \leqn{t5}.  After $SU(2)\times U(1)$ breaking, the $t_L$
and $t_R$ mix with one another, and with a third field, 
the $\chi_t$ in \leqn{t5}. In Appendix D, we construct the propagator
for these three fields and use it to extract the top quark
wavefunctions.  It turns out that the component of this wavefunction
containing $\chi_t$ is of order $s^2$.  In the cross section
calculation, the overlaps between this term and other terms in the
wavefunction have additional suppression by powers of $s$.  So, we can
ignore the $\chi_t$ contribution in this calculation.

The wavefunction of the top quark can then be written in terms of the
left- and right-chirality components of the Dirac fields $t_L$ and
$t_R$.   In the Dirac basis in which $\gamma^5$ is diagonal, write a
massive Dirac spinor with mass $m_t$, spin $s$  and momentum $p$ as 
\beq
         U(p) =    \pmatrix{  u^s_L(p)  \cr   u^s_R(p)\cr  } \ .
\eeqn
Similarly, the spinor of an antifermion is 
\beq 
      V(p) =   \pmatrix{ v^{-s}_L(p)  \cr   v^{-s}_R(p) } \ ,
\eeqn
where $v^s_L(p) = u^s_L(p)$, $v^s_R(p) = - u^s_R(p)$, and 
 $(-s)$ denotes the flipped spin~\cite{PS}.  Then, ignoring
$\chi_t$, working to order $s^2 \sim (m_t z_R)^2$,  and ignoring small
terms from the $\theta_b$ mixing, we show in 
Appendix D  that the physical top quark wavefunction takes the form
\beq
  \ket{t_{phys}} =\pmatrix{  u_L(p) f_L(z) \left[ (1+A(z) \, m_t^2 z_R^2)
      \ket{t_L}
 + B(z) \, m_t z_R \ket{t_R} \right] \cr
u_R(p) f_R(z) \left[ C(z) \, m_t z_R \ket{t_L} + (1+ D(z) \, m_t^2 z_R^2)
  \ket{t_R} \right] \cr}  \ . 
\eeq{twf}
The functions $f_L(z)$ and $f_R(z)$ are the 
 left- and right-handed zero mode wavefunctions defined in
 \leqn{leftzero} and 
\leqn{genrightzero}. The kets in the brackets denote the quantum
numbers of the corresponding Dirac fields. The expressions for  the coefficients
$(A,B,C,D)$ are given in 
Appendix D.  It should be noted that each of the chiral 
wavefunctions is separately normalized.
 For example, for the left-chirality term, this implies
 (for simplicity, setting  $a_t = 0$)
\beq
\int {dz \over (kz)^4} |f_L(z)|^2 \left[(1 + A(z) m_t^2 z_R^2)^2 
+ (B(z) m_t z_R)^2 \right] = 1 \ . 
\eeqn

We can now compute the neutral gauge boson propagator between states of
definite helicity for $e^-$ and $e^+$ in the initial state and states
of definite helicity for $t$ and $\bar t$ in the final state.  
The first step is to take the matrix element of \leqn{approxS} 
between the partial wavefunctions in  brackets in \leqn{twf}.
The operators ($T^{3L}_t, Y_t, T^{3R}_t)$ are the pure numbers
$(\half, {1\over 6}, -\half)$ and $(0, {2\over 3}, 0)$, respectively,
acting on $\ket{t_L}$ and $\ket{t_R}$ in \leqn{twf}, and that $T^{35}$  mediates
between $\ket{t_L}$ and $\ket{t_R}$, 
\beq
          \bra{t_L} T^{35} \ket{t_R} =    \bra{t_R} T^{35} \ket{t_L} =
          \half  \  , 
\eeqn
without changing the chirality. For $t_L$, $\delta Q$ in \leqn{devs}
receives a nonzero contribution from the first term in parentheses
proportional to $s^2$.  For both $t_L$ and $t_R$, there is a nonzero
contribution  from the second term, proportional to  $s\cdot  (m_t
z_R)$, 
where the second factor comes from the
wavefunction overlap between the opposite chirality components.

Finally, we compute the overlaps of the final-state spinors.   For
example,  for $e^-_Le^+_R \to t_L \bar t_R$, where the $t_L$ and $\bar
t_R$
now 
refer to physical top quark and antiquark helicity eigenstates,
\beq 
     v^\dagger_L(e^+) \bar \sigma^m u_L(e^-) \   u^\dagger_L(t) \bar
     \sigma_m v_L(\bar t)  =    2 E (E+p) (1 + \cos\theta)  \ , 
\eeqn
and for  $e^-_Le^+_R \to t_L \bar t_L$, again, for the physical
helicity eigenstates,
\beq 
     v^\dagger_L(e^+) \bar \sigma^m u_L(e^-) \   u^\dagger_L(t) \bar
     \sigma_m v_L(\bar t)  =    - 2 E m_t \sin \theta  \ ,  
\eeqn
where $(E, p)$ are the final energy and momentum of the top quarks. 
Assembling all of the pieces, we find expressions for the
helicity-dependent differential cross sections in the form
\leqn{basiccsfort}.

\subsection{Deviations of the pair-production cross section}

We now present  the numerical values for the deviations of the helicity
cross sections for $\ee\to t\bar t$ from the predictions of the SM.
Since we have seen in Section~5.2 that the case of $t$ in the {\bf 4}
of $SO(5)$ is disfavored by the constraint from $R_b$, we will only
consider here the case of $t$ in the {\bf 5}. 

In our discussion of the $b$ quark cross sections, the term $\delta Q$
in \leqn{devs} 
gave the dominant effect at low energies.    Thus we predicted
relatively small corrections to the SM for $b_R$ in the {\bf 5} where
$\delta Q = 0 $   but
large corrections for $b_R$ in the {\bf 4}.   In the top quark case,  $\delta Q$ is
nonzero, though it has a different origin.  Specifically, 
\beq
\delta Q_{t_L} = \left( - {s^2 \over 2}
 + {s \over \sqrt{2}} T^{35} \right) \VEV{z^2 \over z_R^2} \ ,
\quad \delta Q_{t_R} =  {s \over \sqrt{2}} T^{35} \VEV{z^2 \over z_R^2} \ ,
\eeqn
For $t_L$, the moment $\VEV{z^2/z_R^2}$ is suppressed,  since $t_L$ 
is partially elementary when $0.3 < c_t < 0.6$.   On the other hand,
in the same range of $c_t$, $t_R$ is highly composite, 
leading to a quite significant contribution.

Enhancement of the $t_L\bar t_L$ and $t_R\bar t_R$ cross sections is
the standard signal of an anomalous top quark magnetic moment induced
by new physics~\cite{KLY}.   In this class of models, however, the enhancements
of these cross sections are not especially large.

\begin{figure}
\centering
\includegraphics[width=5in]{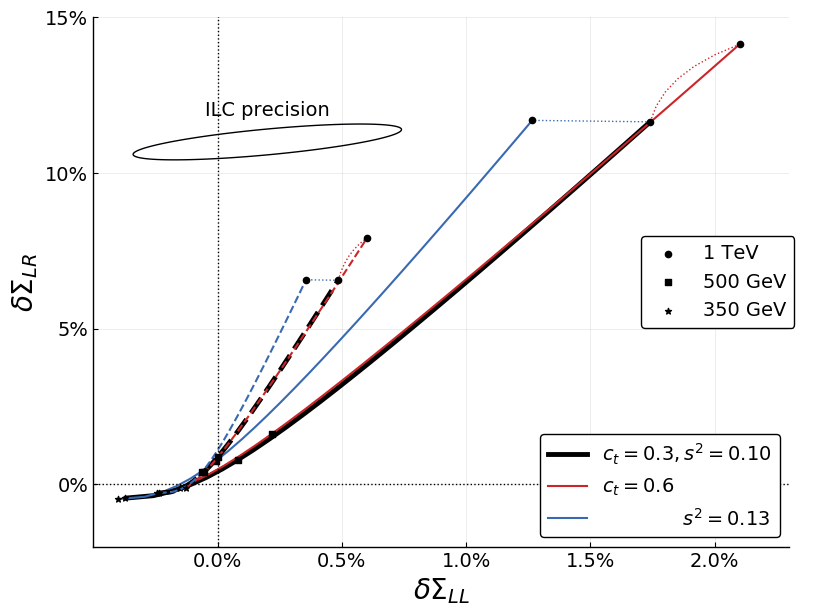}
\includegraphics[width=5in]{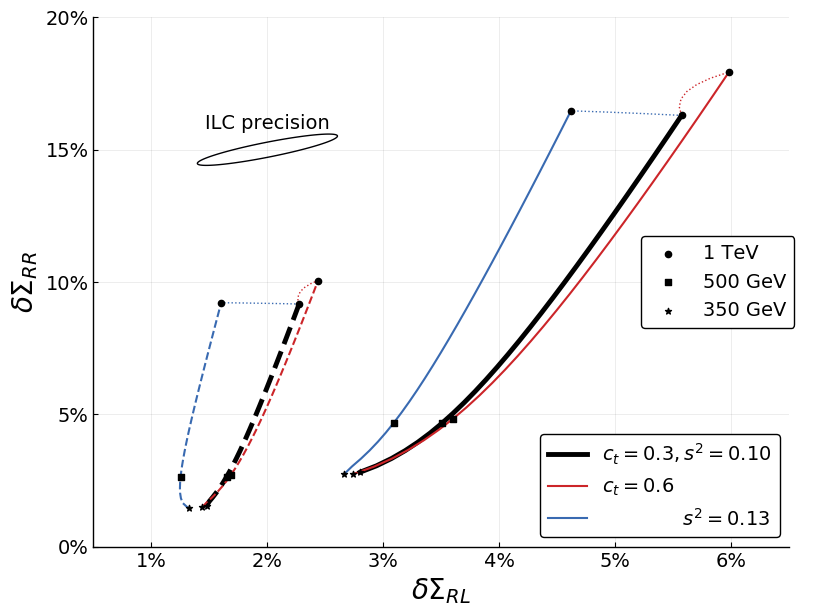}
\caption{Cross section deviations of $\ee \to t_L {\bar t}_R$ and $t_R {\bar t}_L$. $t_L$ and $t_R$ are embedded in {\bf 5} of $SO(5)$.  Solid lines for $k_R = 1.5$~TeV. Dashed lines for $k_R = 2$~TeV.}
\label{fig:t}
\end{figure}

\begin{figure}
\centering
\includegraphics[width=5in]{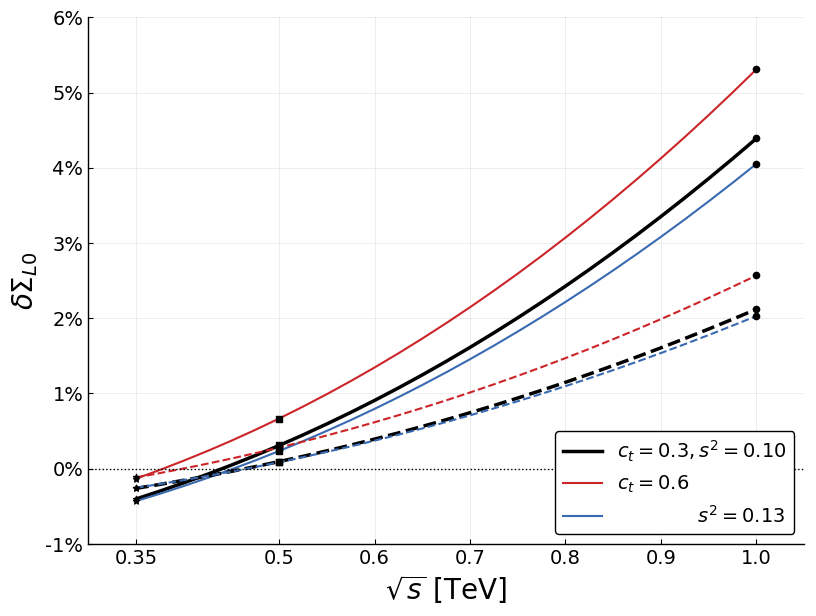}
\includegraphics[width=5in]{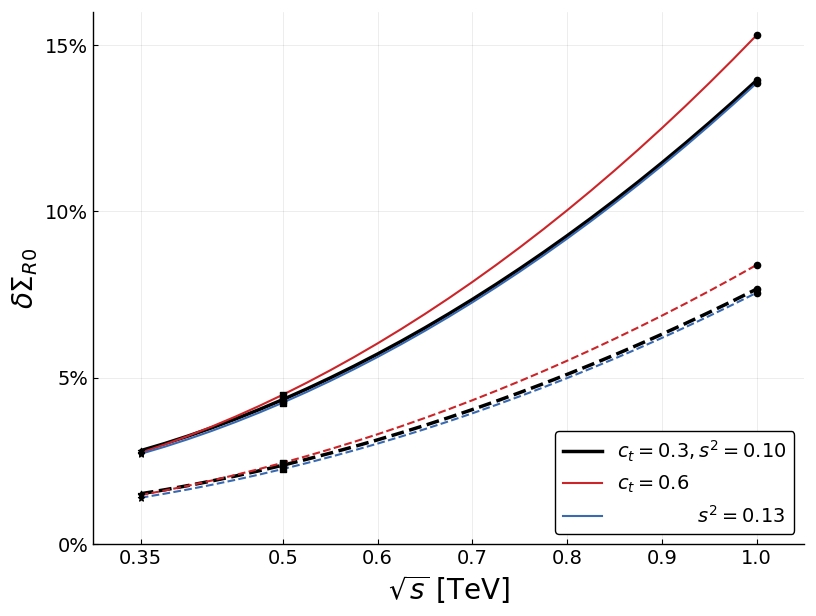}
\caption{Cross section deviations of $\ee \to t_L {\bar t}_L$ or $t_R
  {\bar t}_R$. $t_L$ and $t_R$ are embedded in {\bf 5} of $SO(5)$.
  Solid lines for $k_R = 1.5$~TeV. 
Dashed lines for $k_R = 2$~TeV.}
\label{fig:t0}
\end{figure}

Fig.~\ref{fig:t} shows the cross section deviations for the $t_L\bar
t_R$ and $t_R\bar t_L$ helicity states, in a presentation similar to
that of Fig.~\ref{fig:b5}.    The central values of the parameters are
again taken to be  those in \leqn{centralparams}.   The added lines
show the effect of
 varying the parameters in \leqn{centralparams}
individually,  from 0.3 to 0.6 for $c_t$, from 0.1 to 0.13
for $s^2$, and, finally, with the dashed lines, from $k_R = 1.5$ to
$k_R = 2.0$~TeV. In the two graphs in Fig.~\ref{fig:t}, we 
show our estimate of the 68\% confidence
region for measurements of the two $e^-_L$ cross sections  
and the two $e^-_R$ cross sections that would be obtained at
the International Linear Collider (ILC)  at 500~GeV with 4~ab$^{-1}$
of data.  These estimates are based on results presented in  the
study \cite{Amjad}.  Both the $t_L$ and $t_R$ effects should be
seen clearly for $e^-_R$ beams given the expected accuracy  of the measurements.
Fig.~\ref{fig:t0} shows the cross section deviations  in the
helicity-flip final states $t_L {\bar t}_L$ or $t_R {\bar t}_R$.

\section{Conclusions} 

In this paper, we have present general formulae for the
helicity-dependent cross sections for $\ee\to b\bar b$ and $\ee\to
t\bar t$ in a broad class of RS models with fermions and gauge bosons
in the bulk and $SO(5)\times U(1)$ bulk gauge symmetry.   Through an
analytic understanding of the corrections to the electroweak
propagators induced by the RS structure, we have been able to
understand intuitively when this structure predicts sizable
corrections to the SM expectations that might be visible in precision 
experiments.  Indeed, the precision measurement of $\ee\to b\bar b$
even at 250 GeV in the center of mass can discriminate the models that
we have discussed.  Thus, these measurements can open a new window
into the dynamics of composite Higgs models even at the initial energy
of the ILC.

In the processes we discuss in this paper, the largest effects appear
in subdominant helicity states.   These can be recognized in the $\ee$
cross sections using beam polarization and the high degree of
discrimination between $b$ and $\bar b$ that $\ee$ linear collider
experiments make possible.   We hope that our analysis will be useful
to those who plan for these features in future experiments.

\Acknowledgements

We are grateful to Kaustubh Agashe,  Yutaka Hosotani,
Marcel Vos, and our colleagues in the SLAC Theory
Group for useful discussions and advice.  We are especially grateful
to Roman P\"oschl for  pushing us to closely examine the physics of
$b\bar b$ production at $\ee$ colliders. 
This work was supported by the U.S. Department 
of Energy under contract DE--AC02--76SF00515. 
JY is supported by a Kwanjeong Graduate
Fellowship.

\appendix

\section{Treatment of wavefunctions in the presence of boundary
  kinetic terms}

In our discussion of the fermion zero mode wavefunctions, we saw that
the inclusion of a boundary  kinetic term creates a singular part of the
wavefunction at $z = z_0$ that contains a finite fraction of the
normalization integral.    In this appendix, we describe how we can
compute and 
work with such wavefunctions without needing to deal explicitly with
the singular terms. 

First, we can compute the regular part of the wavefunction from the
propagator.  This is in fact the way that we derived the expression
\leqn{leftzero} and similar expressions for gauge field
wavefunctions in Appendix C of \cite{YPtwo}.   We first compute the Green's functions for
the fields that give rise to the particle, using the rules explained
in that Appendix that account for the boundary kinetic term.   Then we notice that, on
the particle pole, these Green's functions factorize into the form
\beq
     \VEV{\Phi^a(z) \Phi^B(z')} =   { \psi^A(z) \, \psi^{* B} (z') \over
 p^2 - m^2 } 
\eeq{factoriz}
The resulting  wavefuntion $\psi^A(z)$ is correctly normalized acording to a
prescription such as \leqn{leftnorm} that includes the boundary
kinetic term.

This calculation gives only  the smooth part of the wavefunction at $z >
z_0$.  However,  that smooth wavefunction will correctly compute
expectation values of functions that vanish at $z = z_0$.   Since the
wavefunction is normalized when its singular term is included, 
expectation values of general functions can be computed using a generalization of the
formula \leqn{expformforf}.

\section{Neutral gauge field propagator}

In \cite{YPtwo}, we constructed the low-energy limit of the Green's
function for the neutral gauge bosons in the $SO(5)\times U(1)$
model. Here we present an alternative, and somewhat more transparent,
 derviation of the expression
for this Green's function in the limit that we require for this paper.

In Appendix E of \cite{YPtwo}, we presented the following
representation for this Green's function:  For $z_1 < z_2$, 
\beq
\G (z_1, z_2, p) = k z_1 z_2 \, G_{UV}(z_1, z_0) \, ({\bf C}^\dagger)^{-1} \, G_{IR}(z_2, z_R) U_W^\dagger \ , 
\eeq{SO5G}
$G_{UV}$ and $G_{IR}$ are diagonal matrices of $G$ functions which
satisfy the UV and IR boundary conditions, respectively.  The
expression \leqn{SO5G} makes intuitive sense; it satisfies the correct
boundary conditions on each of  the two branes, and it has the correct poles in
$p$, which must appear at the zeros of $\det {\bf  C}(p)$. In the basis
$(A^{3L}, B, Z', A^{35})$ defined in \leqn{BZ} and \leqn{DM}, we
have $G_{IR}$ and $G_{UV}$ take the form
\beq
    G_{UV}(z_1, z_0) =  \pmatrix{  
    G_{W+-}(z_1,z_0) & & & \cr
    & G_{B+-}(z_1,z_0) & & \cr
    & & G_{++}(z_1,z_0) & \cr
    & & & G_{++}(z_1,z_0)}  \   ,
\eeqn
and  
\beq
    G_{IR}(z_2, z_R) =  \pmatrix{  
    G_{+-}(z_2,z_R) & & & \cr
    & G_{+-}(z_2,z_R) & & \cr
    & & G_{+-}(z_2,z_R) & \cr
    & & & G_{++}(z_2,z_R)}  \   ,
\eeqn
where we have  used the abbreviation $G_{W+-}(z_1,z_0) = 
G_{+-}(z_1,z_0) + a_W p z_0 G_{++}(z_1,z_0)$, and similarly for $G_{B+-}(z_1,z_0)$. 

The Wilson line element \leqn{UW}  has the form
\beq
    U_W =  \pmatrix{  
(1+c)/2  & s_\beta (1-c)/2 & c_\beta (1-c)/2  & -s/\sqrt{2} \cr
s_\beta (1-c)/2 & c_\beta^2 + s_\beta^2(1+c)/2 & 
-c_\beta s_\beta (1-c)/2 & s_\beta s/\sqrt{2} \cr 
c_\beta (1-c)/2 & -c_\beta s_\beta (1-c)/2     &
s_\beta^2 + c_\beta^2 (1+c)/2 & c_\beta s/\sqrt{2} \cr
s/\sqrt{2} & -s_\beta s/\sqrt{2} & -c_\beta s/\sqrt{2} & c }  \   , 
\eeq{UinZ}
and the {\bf C} maxtrix is given by ${\bf C}_{AB}  =  U_W^{AB}  G_{-\alpha,-\beta}(z_0,z_R)$, where $\alpha$ and $\beta$ are the UV boundary condition of the $A$ field and the IR boundary conditions of the $B$ field, respectively. Explicitly, 
\beq 
{\bf C} = \pmatrix{  {(1+c)\over 2} G_{W--}
   & s_\beta { (1-c)\over 2} G_{W--} & c_\beta {(1-c)\over 2}  G_{W--} & 
 -{s\over \sqrt{2}} G_{W-+}\cr
s_\beta  {(1-c)\over 2} G_{B--} &
 (c_\beta^2 + s_\beta^2  {(1+c)\over 2}) G_{B--}  & 
- c_\beta s_\beta {(1-c)\over 2} G_{B--}  &   s_\beta {s\over \sqrt{2}}
G_{B-+} \cr 
  c_\beta  { (1-c)\over 2} G_{+-}  &  -c_\beta s_\beta {(1-c )\over 2}
  G_{+-}
 & ( s_\beta^2 + c_\beta^2
{ ( 1+c )\over 2}) G_{+-}      &   c_\beta
{ s\over \sqrt{2}} G_{++}  \cr 
{s\over \sqrt{2}}  G_{+-} & -s_\beta {s\over \sqrt{2}}  G_{+-} & 
-c_\beta {s\over \sqrt{2}}  G_{+-} &  c \ G_{++} \cr } \ .
\eeq{CaZforZ}
The determinant of {\bf C} is 
\beq
  \det {\bf C} = G_{+-} \biggl[   G_{++}G_{W--}G_{B--}  -  {s^2\over 2 p^2 z_0 z_R}(G_{B--}+s_\beta^2 G_{W--}) \biggr] \ . 
\eeqn

To evaluate \leqn{SO5G}, we need the inverse of {\bf C}.   However,
note that, if the initial fermion states are extremely UV-localized, as we
assume for the electrons in $\ee\to f\bar f$, then we will evaluate
$\G (z_1, z_2, p)$
at $z_1 = z_0$.  In this case, the third and fourth rows of $G_{UV}$
involve only $G_{++}(z_0,z_0) = 0$, so they are identically zero.
Then we only need to work out the first two rows of  $({\bf
  C}^\dagger)^{-1}$, or, equivalently, the first two columns of ${\bf
  C}^{-1}$. We find that these can be written in the relatively simple form
\beq
({\bf C}^\dagger)^{-1} =  {G_{+-} \over \det {\bf C} } \left( \C_Q + \C_L + \C_B \right) \ ,
\eeq{decomC}
where
\beqa
\C_Q &=& - {s^2 \over 2 p^2 z_0 z_R}
\pmatrix{ s_\beta^2 & s_\beta & 0 & 0 \cr 
s_\beta & 1 & 0 & 0}
\CR
\C_L &=& G_{B--}
\pmatrix{ {1+c \over 2} G_{++} & 
s_\beta {1-c \over 2} G_{++} & 
c_\beta {1-c \over 2} G_{++} & 
-{s \over \sqrt{2}} G_{+-} \cr 
0 & 0 & 0 & 0}
\CR
\C_B &=& G_{W--}
\pmatrix{ 0 & 0 & 0 & 0 \cr 
s_\beta {1-c \over 2} G_{++} & 
(c_\beta^2 + {1+c \over 2} s_\beta^2) G_{++} & 
-s_\beta c_\beta {1-c \over 2} G_{++} & 
s_\beta {s \over \sqrt{2}} G_{+-}}   \ .
\eeqan

For  the reaction $\ee \to f \bar f$, the RS form factor $S(p)$
defined in \leqn{eeff} and \leqn{fullS}  takes the form
\beq
S(p) = k {\cal Q}_e^T z_0 G_{IR}(z_0,z_0) ({\bf C}^\dagger)^{-1} 
\VEV{ z  G_{UV}(z, z_R) } U_W^\dagger {\cal Q}_f
\eeqn
where
\beq
{\cal Q}_e = \pmatrix{ g_5 T^{3L}_e \cr g_{5Y} Y_e} \, \qquad 
{\cal Q}_f = \pmatrix{g_5 T^{3L}_f \cr g_{5Y} Y_f \cr 
(g_5 /c_\beta) \left( T^{3R}_f - s_\beta^2 Y_f \right) \cr g_5
T^{35}_f}  \ .
\eeqn
We can evaluate this expression using the decomposition 
\leqn{decomC}, to find
\beq
S(p) = S_Q(p) + S_L(p) + S_B(p) \ ,
\eeqn
where
\beqa
S_Q(p) &=& - {g_5^2 k \, s_\beta^2 \, G_{+-} \over p^2 \det {\bf C}} 
\left( s^2 \over 2 p^2 z_0 z_R \right) Q_e Q_f \VEV{ p z \, G_{+-}(z,z_R)} \CR
S_L(p) &=& {g_5^2 k \, G_{+-} G_{++} G_{B--}  \over p^2 \det {\bf C}} \, T^{3L}_e \CR
	& & \hskip -0.3in \times \left[ T^{3L}_f \VEV{ p z \, G_{+-}(z,z_R)} 
	+ \left( {s^2 \over 2} \left(-T^{3L}_f + T^{3R}_f \right) 
+ {sc \over \sqrt{2}} T^{35}_f \right) \VEV{ z \, G_{++}(z_0,z) \over z_R \, G_{++}} \right] \CR
S_B(p) &=&  {g_5^2 k \, s_\beta^2 \, G_{+-} G_{++} G_{W--}  \over p^2 \det {\bf C}} \, Y_e \CR
	& & \hskip -0.3in  \times \left[ Y_f \VEV{ p z \, G_{+-}(z,z_R)} 
 	- \left( {s^2 \over 2} \left(-T^{3L}_f + T^{3R}_f \right) 
 	+ {sc \over \sqrt{2}} T^{35}_f \right) \VEV{ z \, G_{++}(z_0,z) \over z_R \, G_{++}} \right]
\eeqa{decomS}

We can get some  insight into this expression by computing its low
energy approximation.  To do this, we expand the $G$ functions for 
small $s^2$ and $p^2 z_R^2$,
\beqa
G_{++} &=& {z_R \over 2 z_0} \left[ 1 - {p^2 z_R^2 \over 8} \right] \CR
G_{W--} &=& L_W \left[1 - {p^2 z_R^2 \over 4} \left( 1 - {1 \over L_W} \right) \right] \CR
G_{B--} &=& L_B \left[1 - {p^2 z_R^2 \over 4} \left( 1 - {1 \over L_B} \right) \right] \CR
pz G_{+-}(z,z_R) &=& 1 + {p^2 z_R^2 \over 4} \left( -1 + {z^2 \over z_R^2} + 2 { z^2 \over z_R^2}  \log {z_R \over z}  \right)  \CR
{z \, G_{++}(z_0,z) \over z_R \, G_{++}} &=& {z^2 \over z_R^2} \ ,
\eeqa{expansions}
neglecting  terms of order $z_0^2/z_R^2$.

The first zero of $\det {\bf C}$ is identified with the $Z$ boson
mass.  The location of this zero is at $m_Z^2$ such that 
\beq 
m_Z^2 z_R^2 = s^2  \, { L_B + s_\beta^2 L_W\over L_W L_B  }
	\left[ 1+ { m_Z^2 z_R^2 \over 4 } \left( {3 \over 2} 
	- {c_w^2 \over L_W} - {s_w^2 \over L_B} \right) \right] \  ,
\eeqn
where $c_w= \cos\theta_w$ and $s_w = \sin\theta_w$, defined in terms
of model parameters in  \leqn{4Dcouplings}.  Similarly, 
\beq
\det {\bf C} = G_{+-}G_{++}G_{W--}G_{B--} \left( 1 - {m_Z^2 \over p^2} \right) 
	\left[ 1- { m_Z^2 z_R^2 \over 4 } \left( {3 \over 2} 
	- {c_w^2 \over L_W} - {s_w^2 \over L_B} \right) \right]  \ .
\eeq{detexpansion}

It is easiest to begin by checking the leading-order terms of
\leqn{decomS}. 
Applying the leading terms in \leqn{expansions} and
\leqn{detexpansion} to \leqn{decomS} and identifying  $g^2 = {g_5^2 k
  / L_W}$, $e^2 = g^2 \sin^2\theta_w$, 
we find
\beqa
S_Q(p) &=& -{e^2 m_Z^2 \over p^2 (p^2 - m_Z^2) } \, Q_e Q_f \CR
S_L(p) &=& {g^2 \over p^2 - m_Z^2} \, T^{3L}_e \, T^{3L}_f \CR
S_B(p) &=&  {s_w^2 \over c_w^2} {g^2 \over p^2 - m_Z^2}  \, Y_e Y_f  \ .
\eeqa{decomSapprox}
The sum of the three components gives exactly the SM expression for
the sum of the  photon and Z propagators. 

We can recompute $S(p)$ in a similar way 
including the first corrections in \leqn{expansions} 
and \leqn{detexpansion}. After a straighforward calculation, 
we find the expression shown in \leqn{approxS},  with \leqn{devs} and \leqn{contactW}.

\section{Top and bottom quark masses}

In this appendix, we compute the top and bottom quark masses from poles of their Green's functions. The poles are determined by zeros of the determinant of {\bf C} matrix\cite{YPtwo}, which is given by ${\bf C}_{AC}  =  U_b^{AB}U_W^{BC}  G_{-\alpha,-\gamma}(z_0,z_R)$, where $\alpha$ and $\gamma$ are the UV boundary condition of the $A$ field and the IR boundary conditions of the $C$ field, respectively. $U_b$ correpsponds to the UV mixing matrix with angle $\theta_b$ defined in \leqn{tbmixing}.

We first consider the embedding \leqn{t5} in {\bf 5} of $SO(5)$. In the basis $(t_1,t_2, \chi_t,t_R)$, the {\bf C} matrix for the top quark is
\beq 
{\bf C}_{\bf 5} ^t = \pmatrix{  c_b {(1+c)\over 2} G^t_{t--} & -s_b c_b {(1-c)\over 2} G^b_{t--}
   &  c_b {(1-c)\over 2} G^t_{t--} &  -c_b {s\over \sqrt{2}} G^t_{t-+} \cr
   s_b {(1+c) \over 2} G^t_{+-} & c_b G^b_{+-}
   &  s_b {(1-c)\over 2} G^t_{+-} &  -s_b {s\over \sqrt{2}} G^t_{++} \cr
 {(1-c)\over 2} G^t_{+-} &  0 & {(1+c)\over 2} G^t_{+-}  & {s\over \sqrt{2}}  G^t_{++} \cr
{s\over \sqrt{2}}  G^t_{+-} & 0 & -{s\over \sqrt{2}}  G^t_{+-} &   c \ G^t_{++}
\cr } \ ,
\eeq{top5C}
where $s_b = \sin \theta_b$ and $c_b = \cos \theta_b$. $G^t_{\alpha \beta}$ and $G^b_{\alpha \beta}$ are evaluated at $(z_0, z_R)$ with the mass parameter $c_t$ and $c_b$, respectively. We will write $\cos \theta_b$ explicitly at points of possible confusion between the mass parameter $c_t$ and $\cos \theta_b$. The determinant of the ${\bf C}_{\bf 5} ^t$ matrix is given by
\beq
\det {\bf C}_{\bf 5} ^t = G^t_{+-} G^b_{+-}
\left[ G^t_{++} \left( \cos^2 \theta_b \, G^t_{--} 
+ \sin^2 \theta_b \, G^b_{--} {G^t_{+-} \over G^b_{+-}}
+ a_t p z_0 G^t_{+-} \right) - {\cos^2 \theta_b \, s^2 \over 2 p^2 z_0 z_R} \right] \, .
\eeq{dettop5}
In the basis $(b_1,b_2, \chi_b,b_R)$, the {\bf C} matrix for the bottom quark is
\beq 
{\bf C}_{\bf 5}^b = \pmatrix{  c_b G^t_{t--} & -s_b {(1+c)\over 2} G^b_{t--}
   &  -s_b {(1-c)\over 2} G^b_{t--} &  -s_b {s\over \sqrt{2}} G^b_{t-+} \cr
   s_b  G^t_{+-} & c_b{(1+c)\over 2} G^b_{+-}
   &  c_b {(1-c)\over 2} G^b_{+-} &  c_b {s\over \sqrt{2}} G^b_{++} \cr
0 &  {(1-c)\over 2} G^b_{+-} & {(1+c)\over 2} G^b_{+-}  & -{s\over \sqrt{2}}  G^b_{++} \cr
0& - {s\over \sqrt{2}}  G^b_{+-} & {s\over \sqrt{2}}  G^b_{+-} &   c \ G^b_{++}
\cr } \ ,
\eeq{bottom5C}
The determinant of the ${\bf C}_{\bf 5} ^b$ matrix is given by
 \beq
\det {\bf C}_{\bf 5} ^b = G^t_{+-} G^b_{+-}
\left[ G^b_{++} \left( \cos^2 \theta_b \, G^t_{--} {G^b_{+-} \over G^t_{+-}} 
	+ \sin^2 \theta_b \, G^b_{--} + a_t p z_0 G^b_{+-} \right) - {\sin^2 \theta_b \, s^2 \over 2 p^2 z_0 z_R} \right] \, .
\eeq{detbottom5}
The top and bottom quark masses correspond to the first zeros of the two determinant \leqn{dettop5} and \leqn{detbottom5}. To show the explicit leading order expression of the top and bottom mass, it is convenient to define
\beq
\F (c) = { (z_R/z_0)^{1-2c} - 1 \over 1-2c } \ .
\eeq{temp}
Then, we have 
\beqa
m_t^2  &=& {s^2 \over 2 z_0^2 } 
\left( { \cos^2 \theta_b \, \over \F(c_t) \cos^2 \theta_b + \F(c_b) \sin^2 \theta_b + a_t } \right) 
\left( 1 \over \F(-c_t) \right) \CR
m_b^2  &=& {s^2 \over 2 z_0^2 } 
\left( { \sin^2 \theta_b \, \over \F(c_t) \cos^2 \theta_b + \F(c_b) \sin^2 \theta_b + a_t} \right) 
\left( 1 \over \F(-c_b) \right) \, .
\eeqa{mass5}
Then the ratio of the two masses is given by
\beq
{m_b^2 \over m_t^2} = \tan^2 \theta_b \left( \F(-c_t) \over \F(-c_b) \right) = \tan^2 \theta_b \left( {1+2c_b \over 1+2c_t} \right)
	\left( {(z_R/z_0)^{1+2c_t} -1 \over (z_R/z_0)^{1+2c_b} -1} \right) \ ,
\eeq{fullmbmt}
and for $z_0 \ll z_R$ and postivie $c_t$ and $c_b$, it reduces to \leqn{mbmt}. 

Similarly, we can compute the masses of the top and bottom quark in the {\bf 4} of $SO(5)$, defined in \leqn{t4}. The determinants of the ${\bf C_4}^{t,b}$ are given by
\beqa
\det {\bf C}_{\bf 4} ^t &=& G^t_{+-} G^b_{+-}
\left[ G^t_{++} \left( \cos^2 \theta_b \, G^t_{--} 
+ \sin^2 \theta_b \, G^b_{--} {G^t_{+-} \over G^b_{+-}}
+ a_t p z_0 G^t_{+-} \right) - {\cos^2 \theta_b \, s_2^2 \over 2 p^2 z_0 z_R} \right] \CR
\det {\bf C}_{\bf 4} ^b &=& G^t_{+-} G^b_{+-}
\left[ G^b_{++} \left( \cos^2 \theta_b \, G^t_{--} {G^b_{+-} \over G^t_{+-}} 
	+ \sin^2 \theta_b \, G^b_{--} + a_t p z_0 G^b_{+-} \right) - {\sin^2 \theta_b \, s_2^2 \over 2 p^2 z_0 z_R} \right] \, . \CR
\eeqa{det4}
Note that $\det {\bf C_4}$ and $\det {\bf C_5}$ are identical up to the difference in $s^2$ and $s_2^2$.  The top and bottom quark masses in {\bf 4} are
\beqa
m_t^2  &=& {s_2^2 \over z_0^2 } 
\left( { \cos^2 \theta_b \, \over \F(c_t) \cos^2 \theta_b + \F(c_b) \sin^2 \theta_b + a_t } \right) 
\left( 1 \over \F(-c_t) \right) \CR
m_b^2  &=& {s_2^2 \over z_0^2 } 
\left( { \sin^2 \theta_b \, \over \F(c_t) \cos^2 \theta_b + \F(c_b) \sin^2 \theta_b + a_t} \right) 
\left( 1 \over \F(-c_b) \right) \, .
\eeqa{mass4}
The mass ratio is identical to \leqn{fullmbmt}. 

We can proceed similarly for mixed representation case, that is, $b_L$ in {\bf 5} and $b_R$ in {\bf 4} of $SO(5)$. The calculation of $\det {\bf C}$ in this case gives $m_t$ from \leqn{mass5} and $m_b$ from \leqn{mass4}. Then for the mass ratio, we have
\beq
{m_b^2 \over m_t^2} = \half \tan^2 \theta_b \left( {1+2c_b \over 1+2c_t} \right)
	\left( {(z_R/z_0)^{1+2c_t} -1 \over (z_R/z_0)^{1+2c_b} -1} \right) \ ,
\eeq{mixedfullmbmt}
for small $s^2$. 

It is interesting to compute how $\delta Q^{\bf 4}_{b_L}$ in \leqn{dQ4} is related to the top quark mass. Note that we are interested in the range $0.3 < c < 0.7$ and therefore it is a good approximation to ignore terms with $(z_0/z_R)^{1+2c}$ and $(z_0/z_R)^{3-2c}$. Then, the mass of the top quark in {\bf 4} is 
\beq
m_t^2 = {s_2^2 \over z_R^2 } { 1+2c_t \over (1-(z_0/z_R)^{1-2c_t})/(1-2c_t) + a_t (z_0/z_R)^{1-2c_t} } \, ,
\eeqn
where we neglect the small effect of $\theta_b$ mixing. Also, if we compute the expectation value with $f_L(z)$ in \leqn{leftzero} and \leqn{leftnorm}, we have
\beq
\VEV{z^2 \over z_R^2} = { 1/(3-2c_t) \over (1-(z_0/z_R)^{1-2c_t})/(1-2c_t) + a_t (z_0/z_R)^{1-2c_t} }   \, .
\eeq{theVEVevalfora}
Then, for small $s^2$, we have 
\beq
\delta Q^{\bf 4}_{b_L} = {s^2 \over 4} \VEV{ z^2 \over z_R^2 } = { m_t^2 z_R^2 \over (1+2c_t) (3-2c_t) } \, ,
\eeqn
as in \leqn{dQ4L}.   

To obtain $\delta Q^{\bf 4}_R$, we can determine $\VEV{z^2/z_R^2}$
directly from \leqn{theVEVevalfora} by sending $c_t \to -c_t$ and
setting $a_t = 0$.  This gives 
\beq
 \VEV{z^2 \over z_R^2} = { 1+ 2 c_t\over 3+ 2c_t } \ ,
\eeq{theVEVevalforzero}
which leads directly to \leqn{Qfor4Reval}.

\section{The wavefunction of the $t$ quark in the $SO(5)\times U(1)$
  model}

In this Appendix, we construct the on-shell $t$ quark wavefunction. We
compute the propagator of the multiplet $\Psi_t$ in~\leqn{t5}, and
identify $t$ quark wavefunction from the $t$ quark
 pole terms.   Some clarification of the notation is necessary.

The details of computing fermion Green's functions
 in RS spacetime is again given in~\cite{YPtwo}. Here we
 simply quote essential results. The Green's functions of fields
$\psi_L^A,\psi_L^{\dagger B}$ obeying
$\alpha, \beta$ boundary conditions on the IR brane (therefore, in UV gauge) is given by
\beq
\VEV{ \psi^A_L(z) \psi^{\dagger B}_L(z')}
 = (\sigma \cdot p) k^4  p z_R( z z' )^{5/2} \bigg[ {\bf A}^{AB}
G_{+,-\alpha}(z,z_R) G_{+,-\beta}(z',z_R) + \cdots \bigg] \ .
\eeq{topgreen}
The Green's functions $\VEV{ \psi^A_L(z) 
\psi^{\dagger B}_R(z')}$, 
$\VEV{ \psi^A_R(z) \psi^{\dagger B}_L(z')}$, and $\VEV{ \psi^A_R(z)
  \psi^{\dagger B}_R(z')}$  are constructed similarly,
with $ G_{+,-\alpha} \to  G_{-,-\alpha}$ for each $\psi_R$.

The top quark pole $(p^2 - m_t^2)$ is contained in the matrix ${\bf A}^{AB}$. 
After electroweak symmetry breaking, the physical top quark is a mixture of the three fields $t_L$, $\chi_t$, $t_R$ in~\leqn{t5}. We ignore the effect of the UV mixing with the $\Psi_b$ multiplet, since it is always of higher order in our approximation. In the basis $(t_L,\chi_t,t_R)$, the matrix {\bf A} is given by ${\bf A = C^{-1}D}$ where
\beq 
{\bf C} = \pmatrix{  {(1+c)\over 2} G_{t--}
   &  {(1-c)\over 2} G_{t--} &
 -{s\over \sqrt{2}} G_{t-+}\cr
 {(1-c)\over 2} G_{+-} &   {(1+c)\over 2} G_{+-}  & {s\over \sqrt{2}}  G_{++} \cr
{s\over \sqrt{2}}  G_{+-} & -{s\over \sqrt{2}}  G_{+-} &   c \ G_{++}
\cr } \ ,
\eeq{Catfort}
and 
\beq 
{\bf D} = \pmatrix{ - {(1+c)\over 2} G_{t-+}
   & - {(1-c)\over 2} G_{t-+} &
 -{s\over \sqrt{2}} G_{t--}\cr
 -{(1-c)\over 2} G_{++} &  - {(1+c)\over 2} G_{++}  & {s\over \sqrt{2}}  G_{+-} \cr
-{s\over \sqrt{2}}  G_{++} & {s\over \sqrt{2}}  G_{++} &   c \ G_{+-}
\cr } \ .
\eeq{Datfort}
Note that we use $G_{t-\pm}  =  G_{-\pm} + a_t p z_0 G_{+\pm}$.

More explicitly, we can write
\beq
{\bf A} = { \widetilde {\bf A} \over \det {\bf C} } \ ,
\eeqn
 where 
\beq
  \det {\bf C} = G_{+-} \biggl[   G_{++}G_{t--}  -  {s^2\over 2 p^2 z_0 z_R} \biggr] \ , 
\eeq{tdeterminant}
and
\beqa
\widetilde {\bf A}_{11} &=& -G_{++} \left(G_{++}G_{t--} - \left({1+c\over 2} + {s^2 \over 4}\right){1 \over p^2 z_0 z_R}\right) \CR
\widetilde {\bf A}_{22} &=& -G_{++} \left(G_{++}G_{t--} - \left({1-c\over 2} + {s^2 \over 4}\right){1 \over p^2 z_0 z_R}\right) \CR
\widetilde {\bf A}_{33} &=& G_{+-} G_{+-}G_{t--} \CR
\widetilde {\bf A}_{12} &=& G_{++} \left({s^2 \over 4}\right) {1 \over p^2 z_0 z_R} \CR
\widetilde {\bf A}_{13} &=& G_{+-} \left(-{s \over \sqrt{2}} \right) \left({1+c \over 2}\right) {1 \over p^2 z_0 z_R} \CR
\widetilde {\bf A}_{23} &=& G_{+-} \left(-{s \over \sqrt{2}} \right) \left({1-c \over 2}\right) {1 \over p^2 z_0 z_R} 
\eeqan
with $\widetilde {\bf A} = \widetilde {\bf A} ^\dagger$. 

The first zero of $\det {\bf C}$ gives the top quark pole $p^2 = m_t^2$. Then, on the pole, we have an identity
\beq
G_{t--} = {s^2 \over 2 m_t^2 z_0 z_R G_{++} } \ ,
\eeqn
where $G$ functions are evaluated at $p^2 = m_t^2$. Using this identity, one can show that the $\widetilde {\bf A}$ factorizes onto the pole of a single fermion
\beq
\widetilde {\bf A} = \vec{n} \vec{n}^\dagger
\eeqn
where
\beq
\vec{n}^\dagger = \sqrt{G_{++} \over m_t^2 z_0 z_R} 
\left( {1+c \over 2} , \,  {1-c \over 2}, \,  -{s \over \sqrt{2}} {G_{+-} \over G_{++}} \right) \ .
\eeq{topfactor}
From $(\det {\bf C} )|_{p^2 = m_t^2} = 0$, we can write 
\beq
{1 \over \det {\bf C} } = {1 \over G_{+-}G_{++}G_{t--}} \left( {p^2 \over p^2 - m_t^2} \right)
	\left[ 1 + \delta m_t^2 \right]  \, .
\eeq{topdet}
The expression for $\delta m_t^2$ will be given below.
Now we are ready to construct the full wavefunction of the physical $t$ quark. 
From \leqn{topgreen}, \leqn{topfactor}, and \leqn{topdet}, the left-handed chirality wavefunction $(t_{phys})_L$ of the top quark is given by
\beqa
   \ket{(t_{phys})_L}  &=& {k^2 m_t  z^{5/2} \over \left( m_t z_0 G_{+-} G_{t--} \right)^{1/2} }  \left( 1 + \half \delta m_t^2 \right)
    u_L(p) \CR 
 && \times \Biggl\{ \left( 1+c \over 2 \right) G_{+-} (z,z_R) \ket{t_L}
 + \left( 1-c \over 2 \right) G_{+-} (z,z_R) \ket{\chi_t} \CR
& & \hskip 0.8in  - \left( s \over \sqrt{2} \right) {G_{+-} \over G_{++} } \, G_{++} (z,z_R) \ket{t_R}  \Biggr\} \ .
\eeqa{topfull}
where $u_L(p)$ is 2-component projection of a massive spinor onto left-handed chirality. The right-handed chirality wavefunction $(t_{phys})_R$ is given by replacing $u_L(p) \rightarrow u_R(p)$ and $G_{+,\alpha}(z,z_R) \rightarrow G_{-,\alpha}(z,z_R)$. 

We can expand the top quark wavefunction up to the first corrections in $s^2$ or $m_t^2 z_R^2$. 
It will be useful to
adopt a compact notation for the expansions of the $G$ functions.  We
will write
\beq
G_{++}(z,z_R; p ) =   G_{++} (z, z_R;p=0) \bigl[ 1 +  (p z_R)^2  {\cal
  Z}_{++}(z) + \cdots \bigr] \ , 
\eeq{calZdef}
and similarly for the other $G$ functions, putting the appropriate
subscript on the ${\cal Z}$ coefficient.

Note that $\ket{\chi_t}$ component is suppressed by $(1-c)/2 = \O(s^2)$. 
In the pair production process, the top quark wavefunction always appear as its square, and therefore the $\O(s^4)$ contribution of $\ket{\chi_t}$ can be ignored. Then we have
\beqa
   \ket{(t_{phys})_L}  &=& u_L(p) f_L(z) \Biggl\{ \left[ {1+c \over 2} +  \half \delta m_t^2  
   + m_t^2 z_R^2 \left(-\half \Z_{+-}(z_0) - \half \Z_{t--}(z_0) +  \Z_{+-}(z) \right)  \right] \ket{t_L} \CR
& & \hskip 1in  - { s \over \sqrt{2}} \left(1- \left(z \over z_R\right)^{1+2c} \right) \ket{t_R}  \Biggr\} \CR
\ket{(t_{phys})_R}  &=& u_R(p) f_R(z) \Biggl\{ { s \over \sqrt{2}} {(z_0/z_R)^{c-1/2} \over L_t}
\left(1- \left(z / z_R\right)^{1-2c} \over 1- 2c \right) \ket{t_L} \CR
& & \hskip 1in   \left[ 1 +  \half \delta m_t^2  
   + m_t^2 z_R^2 \left(-\half \Z_{++}(z_0) + \half \Z_{+-}(z_0) +  \Z_{-+}(z) \right)  \right] \ket{t_R}  \Biggr\} \, . \CR 
\eeqa{topapproxUV}
where
\beqa
L_t &=&   G_{t--}(z_0,z_R;p=0) = {1\over 2c-1} \bigl[ ({z_R\over
  z_0})^{c-1/2} -  ({z_0\over
  z_R})^{c-1/2} \bigr] +  a_t  ({z_R\over
  z_0})^{c-1/2} \CR
\Z_{++}(z_0) &=& -{1\over 2 (2c+3)} \CR
\Z_{t--}(z_0) &=& -{1\over 2(2c+1) L_t}\biggl({1\over (2c-3)}\biggl\{ \bigl[
 ( {z_R\over z_0})^{c-1/2} +( {z_0\over z_R})^{c-1/2} \bigr]  \CR
 & & \hskip 0.2in - {2\over 2c-1} \bigl[
 ( {z_R\over z_0})^{c-1/2} -( {z_0\over z_R})^{c-1/2} \bigr]\biggr\} + 
 a_t  ( {z_R\over z_0})^{c-1/2} \biggr) \CR
 \Z_{+-}(z) &=& -{1\over 2(2c+1)} \biggl( 1 - {z^2\over z_R^2} \bigl[ 1
+ {2 \over 2c-1} (1 -  ( {z\over z_R})^{2c-1} ) \bigr] \biggr)
\eeqan
and 
\beq
\delta m_t^2 = { m_t^2 z_R^2 } \left( -\Z_{++}(z_0) - \Z_{t--}(z_0) \right) \, .
\eeqn
We have made the above formulae somewhat simpler 
by ignoring factors of
$(z_0/z_R)$ and $(z_0/z_R)^{c + 1/2}$ (but not $(z_0/z_R)^{c - 1/2}$)
for the relevant values  $c > 0.3$. 
Note the difference in sign of $\Z$ coefficients 
compared to those defined in Appendix G of~\cite{YPtwo}, since we are working in the Minkowski space.

The wavefunction in \leqn{topapproxUV} is in the UV gauge. We can obtain the wavefunction in the IR gauge by applying $U_W$. Then, we get the coefficient $(A,B,C,D)$ defined in \leqn{twf}:
\beqa
A(z) \, m_t^2 z_R^2 &=& \half \delta m_t^2  
   + m_t^2 z_R^2 \left(-\half \Z_{+-}(z_0) - \half \Z_{t--}(z_0) +  \Z_{+-}(z) \right) -{s^2 \over 2}\left(z \over z_R\right)^{1+2c} \CR
B(z) \, m_t z_R &=& { s \over \sqrt{2}} \left(z \over z_R\right)^{1+2c} \CR
C(z) \, m_t z_R &=& { s \over \sqrt{2}} 
\left( {(z_0/z_R)^{c-1/2} \over L_t} \left(1- \left(z / z_R\right)^{1-2c} \over 1- 2c \right) -1 \right) \CR
D(z) \, m_t^2 z_R^2 &=&  \half \delta m_t^2  
   + m_t^2 z_R^2 \left(-\half \Z_{++}(z_0) + \half \Z_{+-}(z_0) +  \Z_{-+}(z) \right) \CR
& & \hskip .3in  +  { s^2 \over 2} 
\left( {(z_0/z_R)^{c-1/2} \over L_t} \left(1- \left(z / z_R\right)^{1-2c} \over 1- 2c \right) -1 \right) \, ,
\eeqa{fullABCD}
where $c=c_t$. Note that expansion parameters $(m_tz_R)^2$ and
$s^2$ are formally of the same order, related by \leqn{topmass}.


\begin{thebibliography}{99}


 \bibitem{Csakireview}
  B.~Bellazzini, C.~Cs\'aki and J.~Serra,
  Eur.\ Phys.\ J.\ C {\bf 74}, no. 5, 2766 (2014)
  [arXiv:1401.2457 [hep-ph]].

\bibitem{CGT}
C.~Csaki, C.~Grojean and J.~Terning,
  Rev.\ Mod.\ Phys.\  {\bf 88}, no. 4, 045001 (2016)
  [arXiv:1512.00468 [hep-ph]].
 

\bibitem{YPone}
  J.~Yoon and M.~E.~Peskin,
  Phys.\ Rev.\ D {\bf 96}, 115030 (2017)
  [arXiv:1709.07909 [hep-ph]].


\bibitem{YPtwo}
  J.~Yoon and M.~E.~Peskin,
 arXiv:1810.12352 [hep-ph]].




\bibitem{RS}
 L.~Randall and R.~Sundrum,
  Phys.\ Rev.\ Lett.\  {\bf 83}, 3370 (1999)
  [hep-ph/9905221].



\bibitem{gaugeHiggsone}
L.~J.~Hall, Y.~Nomura and D.~Tucker-Smith,
  Nucl.\ Phys.\ B {\bf 639}, 307 (2002)
  [hep-ph/0107331].

\bibitem{gaugeHiggstwo}
  M.~Kubo, C.~S.~Lim and H.~Yamashita,
  Mod.\ Phys.\ Lett.\ A {\bf 17}, 2249 (2002)
  [hep-ph/0111327].





\bibitem{CNP}
 R.~Contino, Y.~Nomura and A.~Pomarol,
  Nucl.\ Phys.\ B {\bf 671}, 148 (2003)
  [hep-ph/0306259].




\bibitem{ACP} 
  K.~Agashe, R.~Contino and A.~Pomarol,
  Nucl.\ Phys.\ B {\bf 719}, 165 (2005)
  [hep-ph/0412089].

\bibitem{Hosotani}
 S.~Funatsu, H.~Hatanaka, Y.~Hosotani and Y.~Orikasa,
  Phys.\ Lett.\ B {\bf 775}, 297 (2017)
  [arXiv:1705.05282 [hep-ph]].

\bibitem{Richard}
 F.~Richard,
  arXiv:1403.2893 [hep-ph].

\bibitem{deCurtis}
  D.~Barducci, S.~De Curtis, S.~Moretti and G.~M.~Pruna,
  JHEP {\bf 1508}, 127 (2015)
  [arXiv:1504.05407 [hep-ph]].



\bibitem{Gherghetta}
See, for example,  T.~Gherghetta, in {\it Physics of the Large and Small (Proceedings of
   TASI 2009)},
   C. Csaki and S. Dodelson, eds.  (World Scientific, 2011). 
  [arXiv:1008.2570 [hep-ph]].


\bibitem{RandallSchwartz}
  L.~Randall and M.~D.~Schwartz,
  JHEP {\bf 0111}, 003 (2001)
  [hep-th/0108114].


\bibitem{GW}
 W.~D.~Goldberger and M.~B.~Wise,
  Phys.\ Rev.\ D {\bf 60}, 107505 (1999)
  [hep-ph/9907218].


\bibitem{DHR}
 H.~Davoudiasl, J.~L.~Hewett and T.~G.~Rizzo,
  Phys.\ Rev.\ Lett.\  {\bf 84}, 2080 (2000)
  [hep-ph/9909255], 
  Phys.\ Lett.\ B {\bf 473}, 43 (2000)
  [hep-ph/9911262].

\bibitem{GP}
 T.~Gherghetta and A.~Pomarol,
  Nucl.\ Phys.\ B {\bf 586}, 141 (2000)
  [hep-ph/0003129],
  Nucl.\ Phys.\ B {\bf 602}, 3 (2001)
  [hep-ph/0012378].

\bibitem{Grossman}
 Y.~Grossman and M.~Neubert,
  Phys.\ Lett.\ B {\bf 474}, 361 (2000)
  [hep-ph/9912408].


\bibitem{custodial}
P.~Sikivie, L.~Susskind, M.~B.~Voloshin and V.~I.~Zakharov,
  Nucl.\ Phys.\ B {\bf 173}, 189 (1980).

\bibitem{ATLASLambda}
 M.~Aaboud {\it et al.} [ATLAS Collaboration],
  JHEP {\bf 1710}, 182 (2017)
  [arXiv:1707.02424 [hep-ex]].


\bibitem{Zbb}
  K.~Agashe, R.~Contino, L.~Da Rold and A.~Pomarol,
  Phys.\ Lett.\ B {\bf 641}, 62 (2006)
  [hep-ph/0605341].


\bibitem{5rep} 
  R.~Contino, L.~Da Rold and A.~Pomarol,
  Phys.\ Rev.\ D {\bf 75}, 055014 (2007)
  [hep-ph/0612048].


\bibitem{4rep} 
  K.~Agashe and R.~Contino,
  Nucl.\ Phys.\ B {\bf 742}, 59 (2006)
  [hep-ph/0510164].


\bibitem{LEPEWWG}
 S.~Schael {\it et al.} [ALEPH and DELPHI and L3 and OPAL and SLD Collaborations and LEP Electroweak Working Group and SLD Electroweak Group and SLD Heavy Flavour Group],
  Phys.\ Rept.\  {\bf 427}, 257 (2006)
  [hep-ex/0509008].
  
 \bibitem{Poschl}
 S.~Bilokin, R.~P\"oschl and F.~Richard,
  arXiv:1709.04289 [hep-ex],   and R.~P\"oschl, personal communication.

  \bibitem{PS}
M. E. Peskin and D. V. Schroeder, {\it An Introduction to Quantum
  Field Theory}  (Westview Press, 1995),   Chapter 3.

\bibitem{KLY}
 G.~L.~Kane, G.~A.~Ladinsky and C.~P.~Yuan,
  Phys.\ Rev.\ D {\bf 45}, 124 (1992).

\bibitem{Amjad}
 M.~S.~Amjad {\it et al.},
  Eur.\ Phys.\ J.\ C {\bf 75},  512 (2015)
  [arXiv:1505.06020 [hep-ex]].

 

\end{thebibliography}
\end{document}